\documentclass[%
 reprint,
superscriptaddress,
 amsmath,amssymb,
 aip,
]{revtex4-2}

\usepackage[margin=1in]{geometry}
\usepackage{booktabs}
\usepackage{array}
\usepackage{graphicx}
\usepackage{dcolumn}
\usepackage{float}  
\usepackage{bm}
\usepackage{svg}
\usepackage{orcidlink}
\usepackage{ragged2e}
\usepackage{hyperref}
\usepackage{multirow}
\usepackage{mathptmx}
\usepackage{textcomp}
\usepackage[normalem]{ulem}
\usepackage[mathlines,pagewise]{lineno}
\usepackage[justification=justified,singlelinecheck=false]{caption}
\usepackage{rotating}

\begin{document}


\title{In-situ SHG microscopy investigation of the domain-wall-conductivity enhancement procedure in lithium niobate}

\author{Iuliia~Kiseleva\, \orcidlink{0009-0002-5435-056X}}%
    \email{iuliia.kiseleva@tu-dresden.de}
    \affiliation{Institut für Angewandte Physik, Technische Universität Dresden, 01062 Dresden, Germany}
\author{Boris~Koppitz\, \orcidlink{0009-0001-6586-0947}}
    \affiliation{Institut für Angewandte Physik, Technische Universität Dresden, 01062 Dresden, Germany}
\author{Elke~Beyreuther\,\orcidlink{0000-0003-1899-603X}}
    \email{elke.beyreuther@tu-dresden.de}
    \affiliation{Institut für Angewandte Physik, Technische Universität Dresden, 01062 Dresden, Germany}
\author{Matthias~Roeper\,\orcidlink{0009-0008-4779-7383}}
    \affiliation{Institut für Angewandte Physik, Technische Universität Dresden, 01062 Dresden, Germany}
\author{Samuel~D.~Seddon\,\orcidlink{0000-0001-8900-9308}}
    \affiliation{Institut für Angewandte Physik, Technische Universität Dresden, 01062 Dresden, Germany}
\author{Lukas~M.~Eng\,\orcidlink{0000-0002-2484-4158}}
    \affiliation{Institut für Angewandte Physik, Technische Universität Dresden, 01062 Dresden, Germany}
    \affiliation{ct.qmat: Dresden-Würzburg Cluster of Excellence—EXC 2147, Technische Universit\"at Dresden, 01062 Dresden, Germany}
\begin{abstract}
Conductive domain walls (CDWs) in the uniaxial ferroelectric lithium niobate (LiNbO$_3$, LN) have attracted a lot of interest as potential elements in 2D nanoelectronics, due to their orders-of-magnitude larger electronic AC and DC conductivities as compared to the host material. On the way towards generating standardized CDWs into z-cut bulk LN crystals with controllable geometry and electrical properties, we have encountered setbacks recently: Although the first preparation step, i.e., the established UV-light-assisted liquid-electrode poling, reliably creates fully penetrating hexagonal domains with the DWs being aligned almost parallel to the polarization axis, the second step in the DW 'conductivity-enhancement' process through post-growth voltage ramping, resulted in randomly-shaped DWs as reflected in their different current-voltage (I-V) characteristics even after having applied the same process parameters. To clarify this phenomenon, we present here an \textit{in-situ} and time-resolved second-harmonic-generation (SHG) microscopy investigation of DW samples of different sizes, monitoring the DW evolution during that critical voltage ramping, which allowed us to reconstruct the 3D DW shapes both prior to and after the enhancement process. As a result, we are able to map the temporal changes of the local DW inclination angle $\alpha$, and to quantify the DW velocity. As a consequence, we need to re-assess and re-think the origin of the DW conductivity (DWC) in LN: The hitherto assumed simple connection between $\alpha$ and the DWC can not be generalized, since point defects accumulating along DWs act as extra sources for charge carrier trapping/release, significantly contributing to the DW current.

\end{abstract}


\maketitle

\section{\label{sec:intro}Introduction\protect}

Ferroelectric (FE) domain walls (DWs) very well separate regions of uniform dielectric polarization and measure only a few unit cells in width \cite{Catalan2007}. Beyond being mere boundaries, they may host localized phenomena, such as enhanced conductivity \cite{Meier2015}, and exhibit structural flexibility as well as sensitivity to external fields \cite{Evans2020, Li2019, Schultheiss2023}. This enables precise control over DW nucleation, geometry, and positioning within an FE crystal. These unique properties make DWs appealing for both fundamental research and practical applications.

Conducting DWs (CDWs) in lithium niobate ($\mathrm{LiNbO_3}$, LN) are in the focus here, since they stand out due to the host material's chemical, optical, and thermal stability and inertness \cite{Weis1985, Rusing2019, Schultheiss2022}, necessities that all ensure reliability under various conditions. LN itself is an uniaxial model FE crystal that has been extensively studied and is widely applied \cite{Manzo2012}, making it a well-understood foundation for investigating its domain wall conductivity (DWC). DWs in LN were reported to exhibit conductivities up to seven orders of magnitude larger as compared to the bulk material's conductivity \cite{Ratzenberger2024}, with the DW stability and conductivity being maintained over a broad temperature range from liquid-nitrogen temperature up to 400\textdegree C \cite{Werner2017, Kampfe2020, Zahn2024, Wulfmeier2025}, with electron-polaron hopping being the most reasonable transport mechanism at the present state of knowledge. Their prospective tunability and ability to be erased and rewritten, hence, make them strong candidates for applications as nanoscale on-chip interconnects linking functional elements in integrated circuits  \cite{Zhang2022}, random-access memories \cite{Sun2022, Suna2023}, and memristors \cite{Xia2019, Suna2022, Kampfe2020}, offering both high functionality and integration potential.

For growing hexagonal ferroelectric domains along the crystal z-axis into LN bulk crystals, the \emph{UV-assisted liquid-electrode poling} technique (see ref.~\onlinecite{Ratzenberger2024} and references therein for details) has become state-of-the-art. The first step when creating CDWs is the artificial growth of an inverted domain into the host crystal. We will label this first preparation step with 'poling' and abbreviate as 'pol.' throughout this paper. Historically, the corresponding \emph{as-poled} domain walls showed no or only a slightly higher conductivity than the bulk crystal, which, however, could be significantly enhanced by UV illumination (attention: not to be confused with the UV-light during poling!), as reported in early works\cite{Schroeder2012}. With the help of the c-AFM method, a correlation between the local surface-near DW inclination and the local conductance was identified, resulting in the well-known head-to-head (h2h) and tail-to-tail (t2t) scenarios. However, very few relatively inclined regions remain poorly conductive. Based on this inclination-conductance correlation, an origin of larger conductance in lithium niobate DWs was assumed to be band bending at charged sites.

The first major step in the direction of \emph{UV-excitation-independent DW conductivity enhancement} in LN was accomplished by \citet{Godau2017}. The suggested 'post-poling enhancement' procedure involves ramping a negative voltage, applied to a metal plate electrode mounted to the +z-side of the LN crystal or -z side of the preliminarily poled domain, up to 66\% of the coercive field $E_{C}$ (corresponding to 3.3 kV/mm for 5-mol\% MgO-doped LN, which means 660~V for a 200~\textmu m thick z-cut sample). This second preparation step, the post-growth 'enhancement' procedure, will be referred to as 'enhancement' or, in brief, 'enh.' throughout this paper. During such a voltage ramping, the DWs change their geometry (direction) due to local fields, and move inwards with velocities that are proportional to the proximity to -z-side of the previously monodomain crystal \cite{Kirbus2019}, thus acquiring a positive overall inclination $\alpha$ with respect to the crystal's z-axis in average. Local inclination values of the DW are distributed unevenly, ranging from positive h2h to negative t2t DW types. Figure~\ref{fig:enhancementStages} illustrates both the initial domain poling and the voltage-driven DWC enhancement step with the corresponding DW geometries, in particular the area and inclination changes, as it was concluded from several pioneering works \cite{Godau2017, Kirbus2019}.

With the logic of the DWC enhancement process and its theoretical background being seemingly clear and solid, an attempt was made to investigate how reproducible the DW fabrication is for different DW geometries and conductances (evaluated through $\pm$10V~I-V~characteristics), employing the developed preparation protocols \cite{Ratzenberger2024}. Our recent investigation \cite{Ratzenberger2024} with a larger sample set of 60 LN specimens demonstrated that for the majority of cases, a kind of 'implosion' of the DW structure -- a rapid process, characterized by the formation of a large number of spike domains, being randomly shaped and distributed -- takes place. To avoid such a non-controllable outcome, we modified the 'enhancement' procedure \cite{Ratzenberger2024}: the voltage up-ramping was terminated as soon as the DW current measured across the sample reached the value of 1~\textmu A; this type of 'enhancement' is labeled further on as the 'current-limited' method. Notably, controlling the 'enhancement' process by the DWC itself has allowed us to prevent more than 50 $\%$ of all samples from implosion. Nonetheless, even more perplexing was the fact that -- in contradiction to the claims stated in earlier works -- no obvious connection between the geometry of the DW (its shape, inclination, or density/perimeter length) and the I-V~characteristics was observed. Consequently, these results raised many question marks on the applicability of previously suggested theories that try to explain the origin of such an enhanced DWC in LN DWs, as well as on the exact mechanism of that enhancement process. This is exactly the starting point of the present work, aiming at a twofold clarification: 

\begin{figure*}[!htb]
    \centering
    \includegraphics[width=0.8\textwidth]{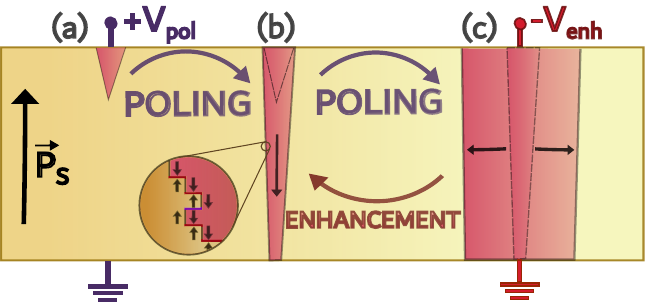}  
    \caption{\justifying
    \label{fig:enhancementStages} Schematic two-dimensional sketch of a z-cut LN crystal with the spontaneous polarization $\vec{P}_S$ illustrating the two fabrication steps of conducting domain walls, i.e., (1) initial domain poling and (2) the 'enhancement' of domain wall conductivity, as they were understood \emph{until now}. During UV-assisted poling, a UV-laser spot (not drawn for simplicity) illuminates the +z surface of the crystal under the application of positive voltage $+V_{pol}$ via liquid electrodes (corresponding electric field: typically 4~kV/mm for 120~s). In the illuminated region, a domain with polarization opposite to the surrounding crystal forms and expands (a), then rapidly propagates through the crystal from the +z to the -z side (b). Subsequently, the hexagonal domain grows laterally while maintaining its shape (c) \cite{Ratzenberger2024}. After metal-electrode deposition on both crystal surfaces, a conductivity enhancement procedure is performed by applying a voltage $-V_{enh}$ to the +z side opposite the polarity of the poling voltage. While the direction of this process is opposite to that of poling, the resulting DW movement differs: the domain contracts, and its walls move inward [stages (c) \textit{back} to (b)]. Unlike during poling, the domain's shape evolves, and the DW velocity varies through the crystal thickness, typically increasing near the -z side, leading to the reformation of \emph{inclined} DWs \cite{Godau2017, Kirbus2019}. In this work, we term the domain wall 'positively inclined' if the structure has more head-to-head than tail-to-tail regions, as depicted in the close-up circle of the domain wall at the stage (b), and 'negatively inclined' otherwise.}
\end{figure*}

First, the question about the role of other factors (apart from DW-inclination) that cause charge accumulation at and contribute to the electrical transport along LN DWs appears. According to the review article by \citet{Nataf2020}, there are \emph{three main origins} of the increased conductance in ferroelectric DWs: 

\begin{itemize}
    \item[(i)] band bending at charged sites (e.g., due to DW inclination) - the mechanism is considered dominant for lithium niobate so far;
    \item[(ii)] band gap reduction at the DW; and
    \item[(iii)] point defect accumulation in the DW vicinity, which acts as a dopant.
\end{itemize}

In particular, by taking a closer look at the motion character of the DWs under an electric field and knowing theoretical information on the mobility of defects as well as their mutual energy stabilization with DWs, as described in detail later in sec.~III.B, we will reassess the origin of the DW conductance in LN as a convolution of at least the mechanisms (i) and (iii) with varying weight, depending on the domain size.

The second motivation is a practical one. Though the UV-assisted liquid-electrode poling step supplies domains of a rather reproducible size and electrical properties, the previous larg-scale study of \citet{Ratzenberger2024} revealed the second step, namely the voltage ramp-up for DW conductivity enhancement, to be a kind of a 'black box' with heavily varying outcomes, as described above. However, an in-depth understanding  of the reasons for this so far unpredictable behavior will be the key to further process optimization with the long-term objective of fabricating DW-based electronics that meet industrial standards.

Consequently, this work comparatively investigates, based on a set of seven samples of different domain sizes, the changes in the geometry for DWs at different layers of the bulk crystal in real-time \emph{throughout the whole DW conductivity enhancement process} using \emph{in-situ} second-harmonic generation-microscopy (SHGM). This approach allows us:
\begin{enumerate}
    \item[(a)] to render the 3D structure of the DWs before and after the voltage treatment to track the changes of the local DW inclination with respect to the crystal's polarization axis in order to review the so-far assumed correlation with the achieved final conductivity (sec.~\ref{subsec:III.A}); and
    \item[(b)] to map the kinetic changes \emph{during} the voltage treatment in the form of local average and instantaneous velocity maps, both derived for the entire perimeter of the DWs, in order to examine the structural evolution and to find the origins behind the seemingly uncontrollable widely varying process outcome (sec.~\ref{subsec:III.B}).
\end{enumerate}

\begin{table*}[t]
\captionsetup{justification=justified, singlelinecheck=false}
\setlength{\tabcolsep}{4pt} 
\renewcommand{\arraystretch}{1.2}
\small 
\begin{tabular}{>{\raggedright\arraybackslash}p{1.5cm} >{\centering\arraybackslash}p{1.5cm} >{\centering\arraybackslash}p{1.7cm} >{\centering\arraybackslash}p{2cm} >{\centering\arraybackslash}p{2cm} >{\centering\arraybackslash}p{2.5cm} >{\centering\arraybackslash}p{1.5cm} >{\centering\arraybackslash}p{1.5cm}}
\toprule
\textbf{Sample} & 
\textbf{Domain Area (µm\textsuperscript{2})} & \textbf{Limiting Current (µA)} & \textbf{Stabilization Voltage (V)} & \textbf{Stabilization Time (sec)} & \textbf{Avg. Inclination Change (deg)} & \textbf{$\mathrm{I_{enh}/I_{as-pol}}$ at -10~V} & \textbf{$\mathrm{I_{enh}/I_{as-pol}}$ at $+$10~V} \\
\hline
A1 & 46720 & 1 & -272 & 1200 & +0.14° & 110 & 86 \\
\hline
A2 & 46255 & 1 & -240 & 300 & +0.01° & 346 & 135 \\
\hline
A3 & 39023 & 1 & -220; -272 & 520; 90 & -0.01°; -0.06° & 195 & 251 \\
\hline
B1 & 20951 & 2 & -248 & 300 & +2.18° & 2.2 $\cdot 10^{6}$ & 2.8 $\cdot 10^{5}$ \\
\hline
B2 & 11735 & 2 & -208 & 300 & +0.34° & 1.4 $\cdot 10^{3}$ & 6.0 $\cdot 10^{5}$ \\
\hline
B3 & 9982 & 2 & -228 & 300 & +1.34° & 4.0 $\cdot 10^{3}$ & 8.4 $\cdot 10^{4}$ \\
\hline
B4 & 19954 & 2 & -228 & 300 & +0.63° & 1.0 $\cdot 10^{3}$ & 1.9 $\cdot 10^{3}$ \\
\bottomrule
\end{tabular}
\normalsize 
\vspace{1pt} 

\caption{\label{tab:1}\justifying Sample data summary: the presented samples were put into two categories by the size of the domains - 'A~group' samples with larger domain of approx. 40'000 \textmu m\textsuperscript{2} and 'B~group' samples with smaller domains of approx. 10'000 -- 20'000 \textmu m\textsuperscript{2}. Besides the domain area, further six key quantities are given: limiting current, stabilization voltage, and time of the 'enhancement' process, as well as the achieved change of the average DW inclinations' absolute values before and after the 'enhancement', as well as the ratios of the DW currents in the before/after state at $\pm$10~V to illustrate the success of the procedure in terms of DW conductivity. Note that sample A3 was subject to a slightly modified comprehensive experimental procedure: there, \emph{two subsequent} 'enhancement' voltage ramps were applied.}
\end{table*}


\section{Materials and Methods}
\label{sec:II}
\subsection{Initial domain growth and sample overview}
\label{sec:II_A}

All seven samples in this study here are cut out as 6 $\times$ 5~mm$^2$ pieces (in x and y crystal directions) from one-and-the same z-cut, 200~\textmu m-thick, 5 mol.\% MgO-doped congruent LiNbO$_3$ wafer (\textit{Yamaju Ceramics Co., LTD.}) to ensure a 100$\%$ stoichiometric compatibility between all experiments.

Creating a hexagonal domain in a single sample was accomplished via the UV-assisted liquid-electrode poling technique as summarized above and sketched in Fig.~\ref{fig:enhancementStages} (see ref. \cite{Ratzenberger2024} and refs. therein for details). Then, 7-8 nm-thick chromium electrodes were thermally evaporated onto both the +z and -z facets of every crystal under high-vacuum conditions (base pressure 10$^{-7}$~mbar). As a result, those electrodes fully covered the formerly generated domains and hence directly connected electrically to the DWs. The relatively thin electrodes were chosen to improve laser transparency for the subsequent DWC enhancement process (see sec.~\ref{sec:II_B}) under \textit{in-situ} SHGM observation. Hence, every sample was individually mounted onto an 80-\textmu m-thin BK7 microscope cover glass, while the top and bottom electrodes were wired to a Keithley 6517B electrometer using conductive silver paste. This allowed 
\begin{itemize}
    \item[(i)] the acquisition of standard $\pm$10-V~I-V~characteristics \emph{before and after} the enhancement process; 
    \item[(ii)] realizing voltage ramp-ups \textit{in-situ} at room temperature up to $\pm$300~V; and
     \item[(iii)] recording \textit{in-situ} SHGM images at any stage of (i) and (ii) through 3D scanning.

    
\end{itemize}

A data summary for all samples is listed in Tab.~\ref{tab:1}. The seven samples were separated into two groups for convenience: Group A samples (A1, A2, A3) all have a single large domain (with an area F~$\gtrsim$~40'000~\textmu m\textsuperscript{2}), while Group B samples (B1, B2, B3, B4) show F~$\lesssim$~20'000~\textmu m\textsuperscript{2} for their domain. As will be shown later, different domain sizes F do severely impact the DWC enhancement procedure.

\subsection{Domain wall-conductivity-enhancement procedure with \textit{in-situ} SHG microscopy observation}
\label{sec:II_B}

\begin{figure*}[htbp]
    \centering
    \includegraphics[width=\textwidth]{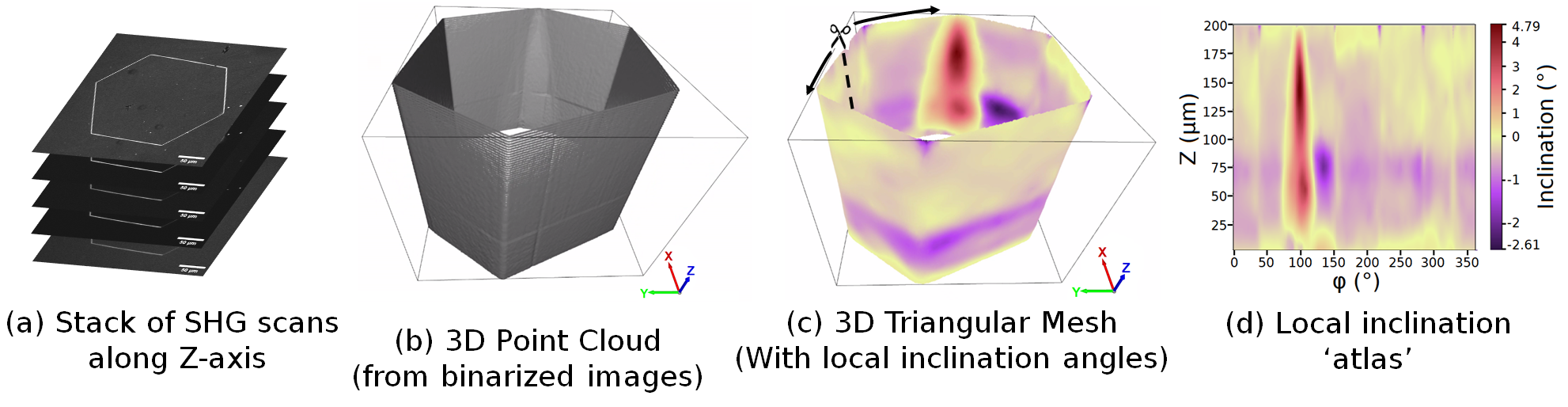} 
    \caption{\justifying
    \label{fig:3DmeshScheme} Stages of the 3D reconstruction of a full SHGM scan of a domain wall along the z-axis. The algorithm, which is described in detail in the supplementary material (SI-secs.~A.1,2) allows the creation of a triangular mesh (c) from the point cloud (b) made out of binarized SHG scans (a). Calculating the local angles between the 3D mesh and z-axis normals allows for an atlas representation (d) of the DW topographical structure.}
\end{figure*}

As motivated in sec.~\ref{sec:intro}, the entire enhancement process is monitored here \textit{in-situ} using SHGM scanning. We use the same SHG setup as reported in ref.~\onlinecite{Kirbus2019}, i.e., the commercial laser scanning microscope Zeiss LSM980MP, and the tunable Ti laser by Spectra Physics (InSight X3, 690–1300 nm, 2 W) with an output laser power of 40 mW at a 900 nm wavelength. The laser generates linearly polarized 100-fs pulses at a repetition rate of 80~MHz. The second harmonic (SH) signal was captured using a focused numerical aperture of 0.8 in the reflection direction.

In LN, the strong second-order nonlinear optical susceptibility $\chi^{(2)}$ enables second-harmonic generation (SHG). FE domains with opposite polarizations exhibit variations in the effective $\chi^{(2)}$ value, especially at DWs where the polarization orientation might change discontinuously. This discontinuity alters the phase-matching conditions in SHG, hence giving rise to the well-localized Cherenkov second-harmonic-generation (CSHG) signal that highlights the DW positions \cite{Rusing2019b}. The SHG image acquisition method is based on raster scanning of a selected area with a chosen resolution. Two different modes of data acquisition were applied here, one with a higher axial image density used always \emph{before and after} the enhancement process (termed 'time-independent' in the following), and one with a lower axial image density, applied \emph{during} the DWC enhancement process (hence termed 'time-dependent' from now on).

For the \emph{time-independent}, static measurements, from which the 3D rendering of the DWs was conducted, approximately one hundred pictures for each domain configuration were captured along the z-axis of the crystal. The lateral (x,y) size of all these scans measured between 250 - 450~\textmu m in width, all recorded with a 1024$\times$1024 pixel resolution. Thus, the axial resolution for the utilized imaging was around 2~\textmu m, while the lateral (x,y) resolution reached the sub-\textmu m level (i.e., $>$ 0.3~\textmu m). 

When performing \emph{time-dependent} SHGM measurements, we need to take into account that the image acquisition velocity is generally slow compared to commercially available optical cameras (in the case of the presented data, the average frame rate was around 2~fps, compared to 30~fps for a typical optical digital camera). This means that with increasing image resolution, the acquisition takes longer, necessitating a trade-off between spatial and temporal resolution of the acquired data. Thus, typically only six layers across the 200-\textmu m-thick sample were recorded. The first and last scanned layers were 25~\textmu m away from the +z and -z surfaces, respectively, with the remaining 4 layers being distributed equidistantly along the z-axes, i.e., separated by $\approx$ 30 \textmu m [see Fig.~\ref{fig:Area}(c)]. In these \emph{time-dependent}, dynamic measurements, the area scanned for the absolute majority of samples was the same as for time-independent domain imaging, however, with the image resolution being reduced to 512$\times$512 pixels. Dynamic SHGM imaging was looped such that after completing the stack of 6 images, the sample position was reset to the first layer z-position close to the -z-facet. Since recording a single frame with the 512$\times$512 pixel resolution takes 0.26~s, every layer is re-allocated after approximately 1.7~s ( = 6 $\times$ 0.26~s). 

The total time of the real-time image acquisition during enhancement-voltage application depends on the individual behavior and dynamics of every sample, a strategy that was motivated by the desire to have as much structural evolution as possible while avoiding any implosion. The corresponding shapes of the enhancement voltage ramps, the accompanying DW current values, the chosen current limits $I_{limit}$, and stabilization times $t_{stab}$, as well as the final voltages $V_{stab}$ can all be found in Tab.~\ref{tab:1} and in the supplementary materials (Figs.~S4 and S7).

To achieve a proper quantitative understanding of both the local geometrical structure changes and the local kinetics of the DWs, the acquired time-resolved gray-scale scans of the samples had to be transformed into morphological skeleton images (binary images consisting of a 1-pixel thick outline of the DWs), which then served as the basis for:
\begin{itemize}
    \item[(a)] rendering a 3D structure of DWs into the form of a 'DW inclination atlas maps' \cite{Wolba2017}, in order to compare the DW states before or after DWC enhancement; and
    \item[(b)] quantifying the DW kinetics under the applied electric field in the form of 'DW-velocity maps', for which the DWs were divided into 360 sectors, each with a 1° width.
\end{itemize}
The main steps of the algorithm are pictured in Fig.~\ref{fig:3DmeshScheme} and described in more detail in the supplementary material (see SI-sec. A.1,2 for the 3D rendering and SI-sec.~A.3 for the DW kinetics analysis).


\begin{figure*}[htbp]
    \centering
    \includegraphics[width=\textwidth]{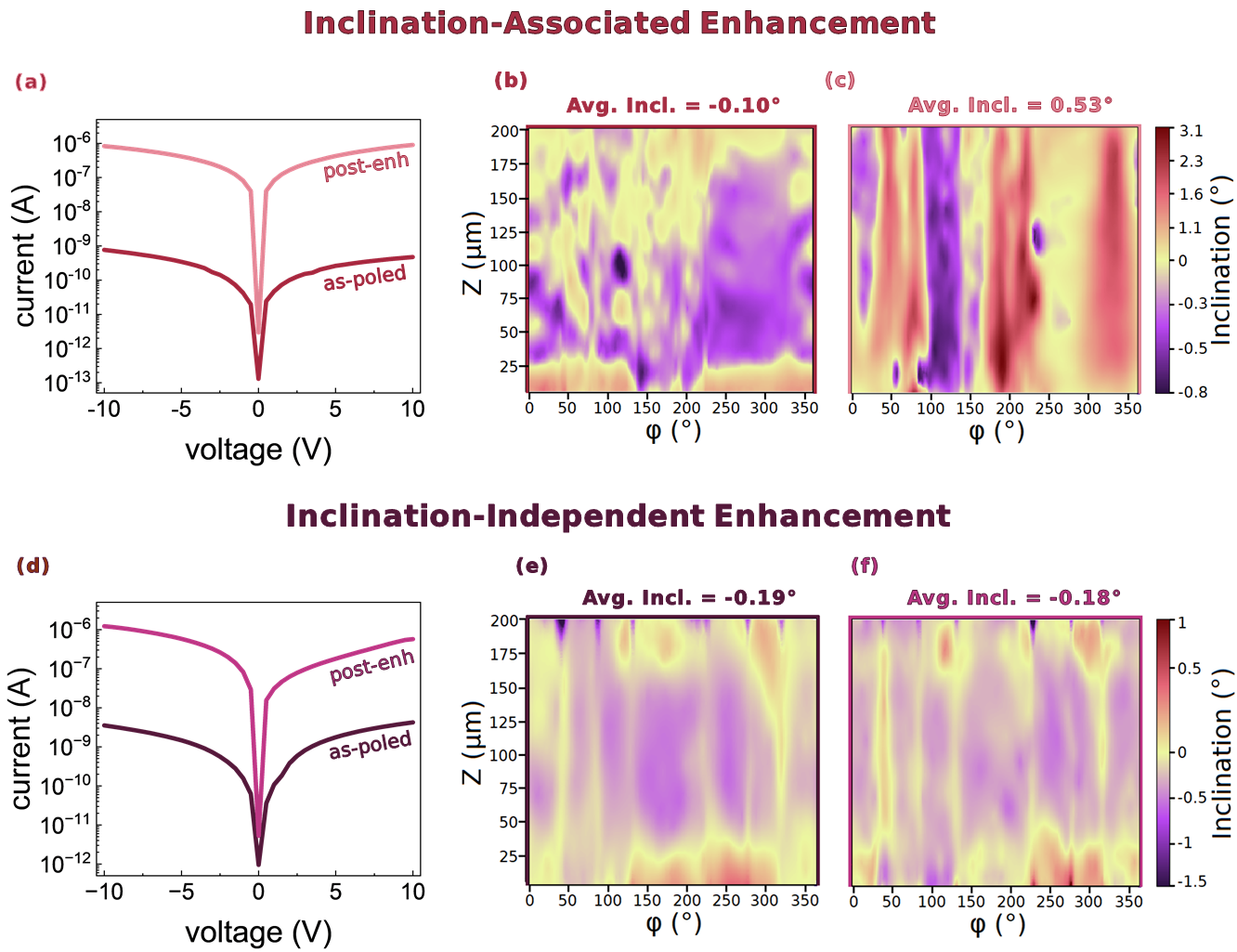}  
    \caption{\justifying \label{fig:InclCond} Comparison of two exemplary DWs' conductance and geometrical evolution through the DW conductivity enhancement by application of voltage ramps (upper row - sample B4; lower row - sample A2). The conductivity changes, showcased in pictures (a) and (d), are very similar for both of them: the increase in conductance is around three orders of magnitude. The geometrical changes, however, are noticeably different - in the case of sample B4, the geometrical structure after the enhancement has acquired large areas with significant positive inclination [transition from (b) to (c)], in full correspondence with results of \citet{Godau2017}. The absence of structural changes for sample A2 [transition from (e) to (f)] is unexpected and implies the existence of a different and/or additional origin for the enhanced DW conductivity, disconnected from the geometrical structure.}
\end{figure*}

\section{Results and Discussion}

In the following, we report -- based on the acquisition of SHGM images and I-V~data of seven LN DW samples divided into the two groups A and B (cf.~sec.~\ref{sec:II_A} and Table~\ref{tab:1}) -- on both 
\begin{itemize}
    \item[(a)] the relationship between the average and local inclinations of DWs and their conductivity, finally leading to a reassessment of previous knowledge \textit{of the dominant conductance mechanism} (sec.~\ref{subsec:III.A}), and
    \item[(b)] the kinetics of DWs under an intermediate electric field (approximately 20~$\%$ of $E_{C}$, being utilized for DWC enhancement), which is qualitatively and quantitatively analyzed (sec.~\ref{subsec:III.B}).
\end{itemize} 
As a result, key factors influencing the DW geometry and conductance -- primarily lattice point defects -- will be identified, along with recommendations for improving the reproducibility in future preparation protocols.

\subsection{Connection between (local) DW inclination and conductance}
\label{subsec:III.A}

From the I-V~characteristics of all DWs under investigation, which are displayed as semi-logarithmic plots and recorded \emph{before} (as-poled) and \emph{after} (post-enh.) the enhancement voltage ramps in Figs.~S5(c,f,j), S8(c,f,i,l) of the supplementary material, we conclude that the procedure was successful (at least 2 orders of magnitude current increase for both polarities as summarized in Table~\ref{tab:1}, last two columns, and in the supplementary material, Fig.~S2) with less variations in the curve asymmetry and general shape than in our previous study~\cite{Ratzenberger2024}. However, when thoroughly evaluating the corresponding 'before/after enhancement' DW inclination atlases, the clear connection between increased DW inclination and increased DW conductivity, which was described in the literature before, does not hold any longer: as a trend, group-A DWs show minimal changes in DW geometry, while group-B DWs display the expected tendency towards positive inclination angles after enhancement. Notably, DWs in both groups A and B experience a significant and equivalent increase in DW conductivity!

To illustrate this key finding in depth, Fig.~\ref{fig:InclCond} shows the I-V~characteristics and 3D inclination atlas structures of two representative samples of each group, samples B4 and A2. Geometrically, the domain area of the first sample is half that of the second, and both samples initially exhibit a predominantly negative average inclination (with prevailing t2t regions). In terms of electrical properties [panels (a) and (d)], the two samples show a similar evolution on a logarithmic scale, with currents increasing from nA to µA at ±10 V bias, while a low-field polarity asymmetry is also visible, more pronounced in A2 than in B4, consistent with different Schottky barriers of the DW/metal interfaces at the z+ and z- sides of the crystal, respectively, as discussed in detail in ref.~\onlinecite{Zahn2024}. However, their geometrical structures evolve in distinctly different ways. The upper panels (b) and (c), referring to B4, depict a structural development as expected from the literature: as described by \citet{Godau2017}, increased conductance correlates with the emergence of extensive positively inclined h2h regions spanning the crystal. These regions serve as primary pathways for charge transport, where mobile electrons compensate for the excess charge at static positive h2h sites. On the other hand, for sample A2 we observe a minimal geometrical evolution only [panels (e) to (f)]. The average inclination remains effectively unchanged, and the geometrical features of the as-poled domain are largely preserved, with only minor modifications after the enhancement procedure. Even with the color scale normalized to the range of inclination angles, 99$\%$ of the structure exhibits angles between -0.6 and 0.3 degrees. No continuous, sample-spanning regions with a high density of positively inclined h2h segments that could serve as a percolation path were observed. These results suggest that the increase of the DW inclination may not be essential for conductance enhancement, pointing instead to the influence of a different mechanism.

\begin{figure*}[htbp]
    \centering
    \includegraphics[width=\textwidth]{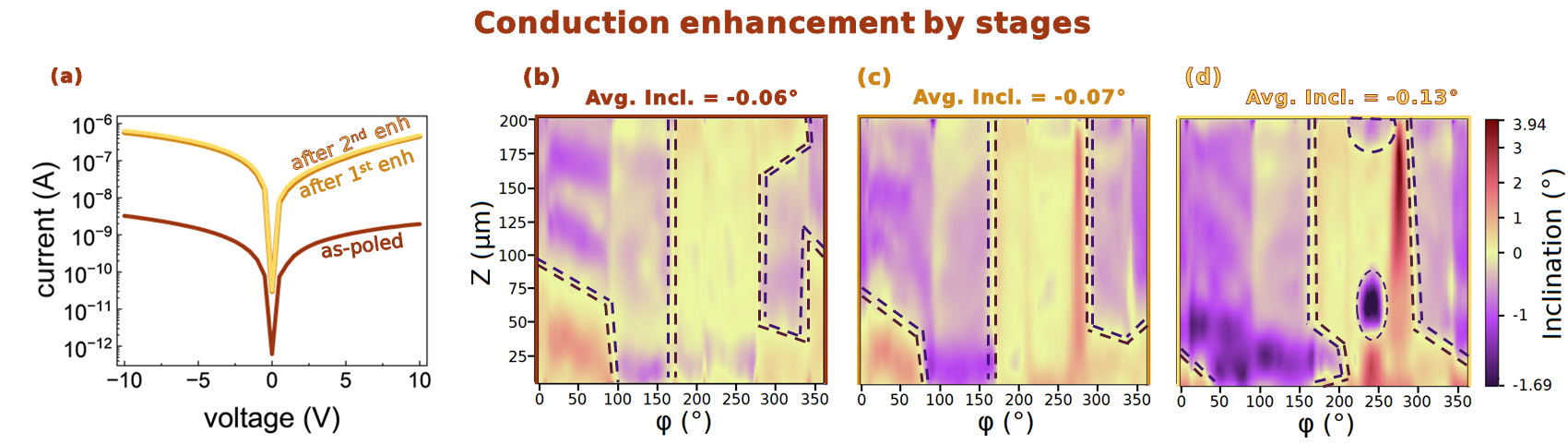}  
    \caption{\justifying \label{fig:Stages} Relationship between DW conductance (a) and geometrical structure at different stages [as-poled (b), after 1st (c) and 2nd (d) enhancement-voltage ramp] for sample A3. As demonstrated here, there are only minute changes of the domain wall inclination observable for the transition from the as-poled state (b) to the state (c) after the first conductance enhancement voltage ramp, but nevertheless, a noticeable increase in conductance of approx 2.5 orders of magnitude is measured. Astonishingly, the second enhancement step led to a significant change in the average inclination but did not cause further changes in the conductance. This result implies that a mechanism different from the formation of h2h/t2t regions plays a decisive role as a source of extra charge carriers -- very probably the accumulation of point defects acting as dopants at the DW.}
\end{figure*}

Further support for this hypothesis comes from the special case of sample A3, which underwent two consecutive enhancement procedures. In Fig.~\ref{fig:Stages}, the evolution of structure and conductance from the as-poled state to the first and second enhancement stages is shown, with the corresponding inclination atlas maps for each stage. The conductance increases only after the first enhancement. Changes between the as-poled state (b) and the post-first-enhancement state (c) are moderate, with some areas migrating and expanding but showing minimal variation in inclination angles, as reflected in the average inclination values. During the second enhancement stage, significant growth in average and local inclinations (d) is observed, yet the conductance remains unchanged, as the almost identical I-V~curves show. This data also strongly suggests the presence of another significant source of conductance in LN DWs, potentially as impactful as band bending at the charged h2h/t2t regions of the DW.

Recapitulating the three mechanisms discussed for DWC above and summarized in the review article by \citet{Nataf2020}, as listed in sec.~\ref{sec:intro}, the most obvious other origin of conductance to consider here is defect accumulation. As demonstrated also by Nataf\textit{ et al.} \cite{Nataf2016, Nataf_2016_2}, there is a bidirectional relationship between DWs and point defects. Defects stabilize DW structures by influencing their local polarization and charge distribution, while the unique DW environment enhances the stability of polar and electronic defect states. Furthermore, defects contribute to DW position stability. For example, in sample A2 [Fig.~\ref{fig:InclCond}(e,f)], the DW may be 'anchored' by numerous defect sites, primarily $Mg_{Li}$ antisites, since they are the ones with the larger calculated stabilization energies\cite{Lee2016}, inhibiting DW motion under applied electric fields due to energetic stabilization. Furthermore, the hypothesis that point defects play a more pronounced role than assumed before is supported by recent results by \citet{Eggestad2024}, who used \textit{ab initio} methods to reveal a strong affinity between uncharged DWs in LN and crystal point defects, such as $V_{Li}$ and $V_{O}$. These defects tend to accumulate near the DW, attracting mobile holes or electrons due to their charge. Combined with the narrower bandgap of neutral DWs compared to bulk LN, this results in a significantly increased conductance. As predicted by Eggestad et al., even a minimal presence of h2h sites within the as-poled DW, coupled to vacancies, leads to a significantly enhanced electronic conduction. Thus, the enhancement procedure’s role in this case, here represented by the group-A samples, would not be the geometrical manipulation of DWs, but rather promoting point-defect accumulation at the DWs -- a phenomenon which has been shown to occur also in other complex oxides \cite{Meng2020}.

Nonetheless, our findings for the group-B DWs (for completeness, their I-V~characteristics and inclination atlases before and after the enhancement procedure are shown in the supplementary material, Fig.~S8) are in complete agreement with previous findings \cite{Godau2017, Zahn2024, Ratzenberger2024}, in that a increased DW inclination goes along with an enhanced conductivity. However, we have to stress that \emph{both} conduction origins, i.e., h2h/t2t regions with possible strong band bending below the Fermi level \emph{and} point defect sites acting as dopants, should not be seen as disjunctive options, but they may rather coexist and the relative contribution to the overall supply of charge carriers confined to the DW might be dependent on the domain size, as these very different trends for samples from the size groups A and B show.

To achieve a deeper insight into the depth-dependent evolution of the DWs under the application of the enhancement voltage, which apparently produces very different final DW structures, being, however, orders-of-magnitude better conductive than the as-grown DWs in \emph{any} case, we proceed with a detailed analysis of the local DW movement in order to address practical challenges on the way towards tunability and reproducibility.

\subsection{Domain wall kinetics under application of large electric fields}
\label{subsec:III.B}

\begin{figure*}[htbp]
    \centering
    \includegraphics[width=0.9\textwidth]{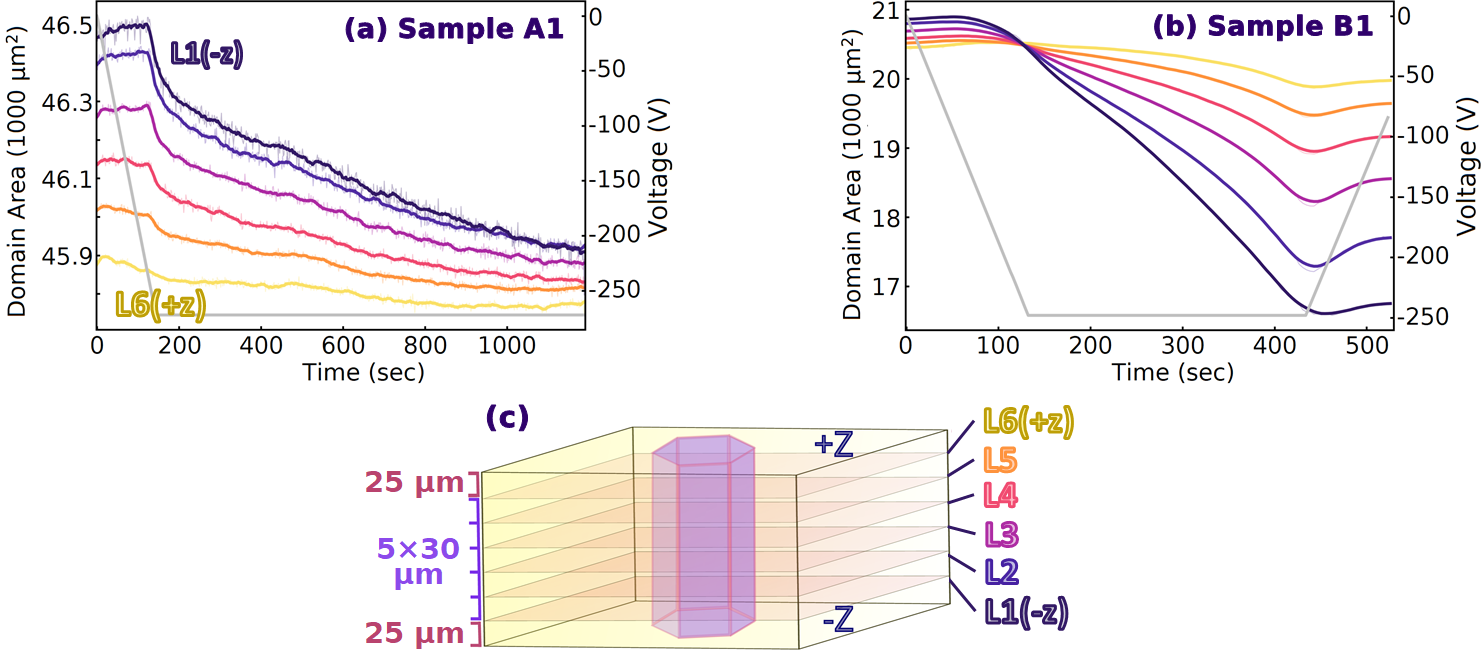}  
    \caption{\justifying
    \label{fig:Area} Domain area evolution during the enhancement process for samples A1 (a), B1 (b), and schematics of layers (c), at which domain wall dynamics was recorded for six equidistantly spaced depths; the dark-blue line corresponds to the layer 30~\textmu m away from the -z-side of the crystal, all the other layers are positioned with steps of 25~\textmu m deeper in the crystal with the yellow line corresponding to the layer 30~\textmu m away from the +z-side, while the gray line shows the applied enhancement/stabilization voltage. For both exemplary samples shown here, as-poled domains are wider towards the -z-side of a crystal, which compares well to the negative average inclination angles of as-poled domains. The kinetic behavior of the domain in sample A1, i.e., the insignificant area change with preservation of the negative average inclination angle, is characteristic for larger domains [the data for the other group A cases, A2 and A3, are depicted in the supplementary material, Fig.~S9(a-c)]. The behavior of sample B1 (mobile, ending up with positive average DW inclination angles) is characteristic for smaller samples [B2, B3, B4, see supplementary material, Fig.~S9(d-f)]. Note that for this sample B1, domain wall 'relaxation' was additionally recorded during the retraction of the applied voltage.} 
\end{figure*}

 In the past, only a few theoretical and experimental studies exploring the DW kinetics in LN and focusing on optical methods \cite{Shur2006, Choi2012, Kirbus2019} have been accomplished, while a \emph{quantitative, systematic and comparative} study of the DW kinetics across different bulk layers in different samples on a macroscopic surface scale has been conducted here.

\subsubsection{Evolution of the domain area as a function of depth and electric field}

In this subsection, we focus on the evolution of the domain area throughout the bulk of the crystal during the enhancement-voltage ramping and the subsequent stabilization period to form an idea of the general mobility of the DW structures and their dependence on position in the bulk. Figure~\ref{fig:Area} illustrates the evolution of areas across layers, starting from the layer positioned 25~\textmu m from the -z-side [indicated by the dark blue line, referred to as L1(-z)], to the layer 25~\textmu m from the +z-side [yellow line, referred to as L6(+z)], with intermediate layers spaced at 30~\textmu m increments. As discussed before, all seven investigated samples can be separated into two (domain-size dependent) groups A and B with markedly different kinetic behavior; here, one representative for each group is shown compared to each other (for the other five samples the respective graphs can be found in supplementary material, Figs.~S9). 

\paragraph{Group A samples -- the case of larger domains: weak DW movement}~

These domains are characterized by their \emph{kinetically hindered behavior} without significant change in the geometrical domain structures. Sample A1 is a typical representative of this group [Fig.~\ref{fig:Area}(a)]: despite a somewhat rapid step towards the area decreasing initially, further domain shrinkage has a very gradual asymptotic character; for layer L1(z-) -- the layer with the highest change in domain area -- the variation in size is only 1.2~$\%$ from its initial value. The 'flip' in the domain area between layers L1(z-) and L6(z+), which is typical for the smaller B-group domains, does not occur. On average, the DW inclination decreased (as confirmed by the 3D structures, cf. sec.~\ref{subsec:III.A}). It is noteworthy that this weak geometrical evolution of the A1 DW is observed despite the largest stabilization time (15~min) and the highest applied voltage (-272~V, -1.36~kV/mm) of all seven samples.

\paragraph{Group B samples -- the case of smaller domains: strong DW movement}~

These domains are characterized by their \emph{kinetically active behavior}. In Fig.~\ref{fig:Area}(b), the group B domains are represented by sample B1, which is the sample with the most mobile DWs among all samples of the study, showing a decrease of its domain area in the layer L1(z-) of 19.7~$\%$. The trend demonstrated by this sample (and also by B2, B3, and B4, as shown in the supplementary material, Fig.~S9) is the following: (i) convergence of domain area values across layers towards uniformity, followed by (ii) a progressive decrease in domain areas with decreasing proximity to the -z-side. This layered behavior suggests a gradient in DW mobility, likely influenced by the applied electric field and local defect concentration, as previous results \cite{Kirbus2019} have shown equally. For sample B1, the relaxation of the DWs under decreasing voltage was also recorded -- what is interesting to note at this stage (Fig.~\ref{fig:Area}(c), after 430~s), the layers L1(-z) and L6(+z) demonstrate the most conservative slopes here as compared to the other layers - layer L2 with the most rapid roll back in terms of area, followed by layers L3, L4 and L5. This fact implies that the crystal surface and/or its interface with the metal electrode serve as a 'stabilizer' for the DWs' inclined geometry, possibly acting as a free-charge supplier for compensation of excessive static charge on the inclined DW.

\begin{figure*}[htbp]
    \centering
    \includegraphics[width=\textwidth]{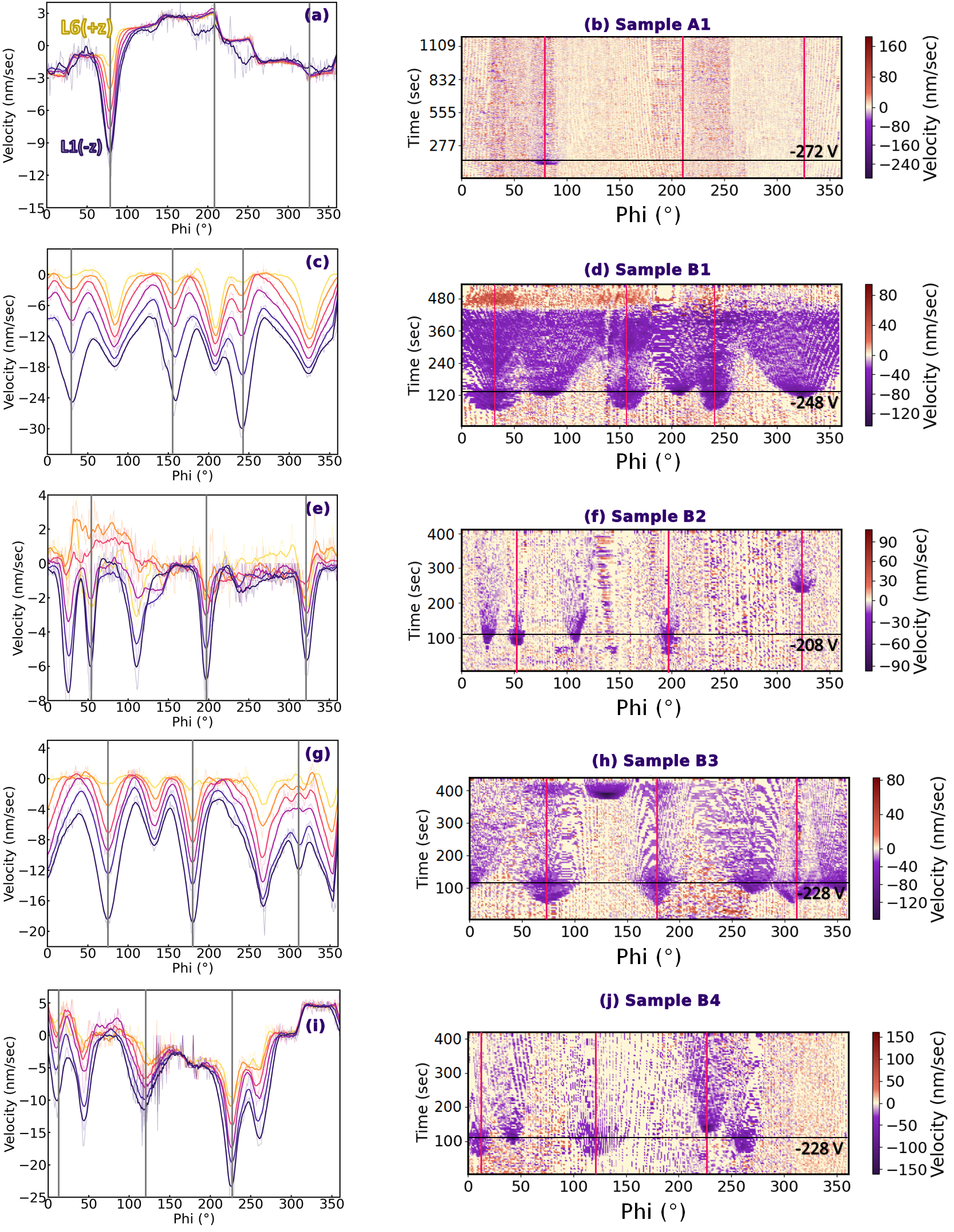}  
    \caption{\justifying
    \label{fig:Local_Average_velocity} Kinetics of domain walls under the application of a negative 'enhancement' voltage to the -z side of the domain, exemplarily shown for samples A1, B1, B2, B3, and B4. The left column diagrams depict the \emph{average local velocity} for each $\phi$-sector of a domain wall. The right column diagrams are a visual presentation of the DW local velocity (color-coded) for each $\phi$-sector of the respective DW \emph{during each time point} of enhancement procedure, shown for layer L1. The vertical lines on the plots demarcate DW sectors that are perpendicular to the y-axis of the LN crystal lattice, specifically the $+$Y corners.}   
\end{figure*}

\subsubsection{Relationship between domain area and the DW mobility -- explanation by pinning phenomena}

As was already hypothesized from the results in sec.~\ref{subsec:III.A}, there is an inverse proportionality between the size of a domain and the mobility of its DWs. It is reasonable to assume that the pinning of the DWs is the defining factor behind their movement, i.e., once a certain ratio between the area of the domain and the local concentration of defects in a crystal is reached, the respective DW becomes immobile. This is indeed one of the scenarios predicted by \citet{Lee2016}: in 5-mol$\%$ MgO-doped LN, the main pinning sites for DWs are $Mg_{Li}$ and despite their much lower energy migration barrier (as compared to $Nb_{Li}$ in congruent LN), the value is still quite high (2.77~eV). For the case of strong pinning of DWs by defects, their motion may proceed between stable positions in the crystal defined by defects.

\subsubsection{Evaluation of the SHGM data in terms of average and time-resolved local DW velocity maps}

Although the walls of the smaller (B group) domains exhibit comparable kinetic behaviors in the bulk of the crystal, both qualitatively (crystal orientation of the domains; patterns of movement) and quantitatively (for samples B1, B2, B3, B4 changes of the domain area in layer L1(z-) are equal to 19.3$\%$, 2.7$\%$, 14.3$\%$ and 5.4$\%$, respectively), they also demonstrate significant discrepancies, which have to be discussed. In order to obtain some detailed insight into the nature of the DW motion, a more thorough evaluation of SHGM data of local DW kinetics is accomplished, ideally with reference to LN crystal axes. Due to the high mobility of the B-group samples' DWs, only they are suitable for such studies; the signal of the group-A samples is largely lost in the noise of the acquisition and processing methods (nevertheless, we also use sample-A1 data for comparison). 

In the left column of Fig.~\ref{fig:Local_Average_velocity}, the \emph{average local velocity} for each DW sector in samples A1, B1, B2, and B3 is shown; six lines represent the six layers, from L1(z-), dark blue, to L6(z+), yellow. The right-column panels present the instantaneous velocity for each 1-degree sector of the DW in layer L1(z-), color-coded and normalized to its minimum and maximum velocity values. Vertical lines on both graphs mark wall sectors lying on the +Y-axis of the LN crystal lattice (see the middle sketch). Following established terminology, sections of DWs parallel to a specific crystal direction are named after that direction \cite{Lee2010}.

An as-poled domain typically forms a hexagonal shape [see supplementary material, Figs.~S3(a,c,e), S6(a,c,e,g)], with each wall being aligned to the Y-wall due to its lower formation energy. While expanding under electric fields and UV irradiation, these Y-walls move as a single front without significant deformation, maintaining the hexagonal shape throughout the poling process \cite{Shur2006, Choi2012}. During the enhancement procedure, however, the walls' movement differs, as it involves reshaping. X-walls, which are the corners of the initial hexagonal domain, begin to move inward, dragging adjacent Y-walls behind, thus changing their direction. According to DFT calculations by \citet{Lee2010}, while Y-walls have lower formation energy, X-walls possess a lower energy barrier for motion. Consistent with previous studies\cite{Godau2017}, our data (both average (left column) and instantaneous (right column) local velocity for all samples, as presented in (Fig.~\ref{fig:Local_Average_velocity}) mostly confirms the previous findings, i.e., the movement in the majority of the cases initiates at the small corners of a DW lying on +Y axes (X-walls), which advance toward the domain center, perpendicular to the crystallographic Y direction. There are, however, some exceptions to this behavior: walls lying in the -Y direction turn out to be more mobile than it was previously reported, and some asymmetry in motion is observed, as pointed out in detail in the following.

\paragraph{The case of $+$Y corners}~

To discuss these discrepancies, a comparison of the movement by sectors [Fig.~\ref{fig:Local_Average_velocity} (b), (d), (f), (h), (j)] is necessary. From there, we can see that the movement of DW sectors lying in +Y corners on average has lower activation. Quantitatively, the beginning of the movement corresponds to the following electric fields: B1 - 0.52~kV/mm; B2 - 0.48~kV/mm (sector at 200°); B3 - 0.44~kV/mm; B4 - 0.48~kV/mm; A1 - 1.0~kV/mm; while for samples A2 and A3 the signal-to-noise ratio does not allow to pinpoint the beginning of the movement. Considering quite a significant light noise and low time resolution of the SHG method, the B-group values can be regarded as equal, being around half of the corresponding value for A1. This observed trend is a new finding, suggesting that larger domain walls in bulk lithium niobate have higher coercive electric fields, making them less mobile.

\paragraph{The case of $-$Y corners}~

Movement of the -Y corners starts at slightly higher fields, an effect which is best visible from movement maps of samples B1 [Fig.~\ref{fig:Local_Average_velocity}(d)] and B3 [Fig.~\ref{fig:Local_Average_velocity}(h)]. For the former the value is equal to 1.0~kV/mm, for the latter to 0.76~kV/mm; for -Y-directed corners of samples B2 and B4 [Fig.~\ref{fig:Local_Average_velocity}(f) and Fig.~\ref{fig:Local_Average_velocity}(j)] These values are 0.86 and 0.89~kV/mm, respectively. Thus, the observation made by \citet{Kirbus2019} concerning the higher mobility of +Y corners compared to -Y, is confirmed. A quantitative difference in velocities, however, turns out to be lower than the previously observed factor-of-4 difference: we state almost no difference in sample B2 [Fig.~\ref{fig:Local_Average_velocity}(e)] or an approximately factor-of-2 difference between the fastest +Y corner and slowest -Y corner in sample B3 [Fig.~\ref{fig:Local_Average_velocity}(c)]. In summary, with a coercive electric field of $E_{C}$=5.5~kV/mm \cite{Kirbus2019} for Mg-doped LN, the initiation of the DW motion lies in the range between 8~$\%$ and 18~$\%$ of $E_{C}$, which also is in good agreement with the previous publication \cite{Kirbus2019}. 

\paragraph{Asymmetries for equivalent corners in one and the same sample}~

Besides the differences in the activation voltages for DW-movement initiation, asymmetries for crystallographically equivalent DW corners can be seen. For example, the corners of A1 at 210° and 325° [Fig.~\ref{fig:Local_Average_velocity}(a,b)] have not changed their position even after approximately 15 minutes under a strong electric field of 1.36~kV/mm, unlike their equivalent corner at 80°. As another example, sample B3 [Fig.~\ref{fig:Local_Average_velocity}(h)] exhibits a significantly delayed movement at the 130° sector (as compared to its crystallographically identical sectors at 0° and 265° -- -Y corners, not marked with pink lines), followed by the rapid jump after 300~s under the electric field, which seems to represent the same phenomenon. Its instantaneous velocity at this time point of 370~s, is noticeably larger than the velocity values of the other corners, and the moving section of the wall has a much broader front than the other corners. There is also a delayed motion after an exposure period to the constant electric field shown by the corner at 325° of sample B2 [Fig.~\ref{fig:Local_Average_velocity}(f)] and the corner at 330° of sample B4 [Fig.~\ref{fig:Local_Average_velocity}(j)]. Overall, it is reasonable to assume that the movement delay (as well as an absence of movement) after the onset of the constant stabilization field is caused by defect migration, as it can be accomplished, e.g., by $Mg_{Li}$ point defects or its clusters, which provide the largest energy stabilization to the DWs in Mg-doped LN \cite{Lee2016}. The expectation time between the movement of all the -Y-corners and the 130° sector of sample B3 [Fig.~\ref{fig:Local_Average_velocity}(h)] of around 270~s would be a defect migration time. 

\paragraph{DW velocity variances and different stabilization voltages within the same size group}~

Another peculiarity arising from the DW velocity maps comes up when we compare the strikingly different 'manner' of DW movement between samples B1 and B2. We see that along the +Y axes, the DWs of sample B1 [Fig.~\ref{fig:Local_Average_velocity}(c)] exhibit an average velocity at least three times higher than that of B2 [Fig.~\ref{fig:Local_Average_velocity}(e)].  If we analyze the velocity color maps of the two samples [Fig.~\ref{fig:Local_Average_velocity}(d,f)], several reasons can be deduced: the instantaneous velocity of the sample B1 domain wall is larger, and the DW movement takes place continuously throughout the whole process, unlike sectors of sample B2 DW, where movement dies out during the process. Additionally, the front of the moving DW in sample B1 is constantly expanding throughout the whole stabilization process, and at the time point of 300~s the whole perimeter of the DW moves inwards. The B2 DW, on the contrary, after an initial movement at its corners, comparatively insignificantly widens its mobile front and the front's expansion dies out together with the movement. At this point we have to stress that also the magnitude of the applied electric field in the enhancement process, which is necessary to achieve the current limit (of 1~\textmu A in the case of group B samples) differs significantly: for sample B2 the voltage ramp-up had to be terminated at a stabilization voltage of -208~V, for sample B1 at -248~V, for instance. In the case of the defect-limited DW motion, it is very likely that, in the case of sample B2, the DW finds the next defect-defined stable position and just gets stuck there. In the case of the higher stabilization voltage of B1, which probably goes along with lower (local) defect concentration, the higher electric field allows both for faster defect migration and also higher DW velocity. Different stabilization voltages which are obviously \emph{individual} for each sample seem to be influenced by the \emph{unique distribution of defects in the vicinity of the DW}, in particular the distribution (and concentration) of (i) the pinning antisite defects $Nb_{Li}$ and $Mg_{Li}$ that delay the geometrical change/inclination acquisition, and of (ii) the vacancy defects $V_{Li}$ and $V_{O}$ playing the role of dopants and making a contribution to conductance.

\subsection{Wrap-up: Decisive role of point defects}

In summary, all evidence discussed above points in the direction of the individual distribution of point defects as being the main factor that defines the exact behavior of an LN DW under the enhancement voltage, with some trend that smaller DWs (due to their smaller surface area) have statistically lower chances to get pinned at one specific position. Considering the fact that all DWs of this study were fabricated from the same nominally homogeneous LN wafer, already very low local concentration variations of point defects may show maximal impact on the electrical DW behavior. Such conclusions match previously reported results on the large role of randomly distributed point defects in other ferroelectrics and their negative effect on the reliable functioning of devices\cite{Yang2024}. For further improvement of the fabrication process of conductive DWs, systematic defect engineering might be the key to enhanced process control and reliability; such attempts were already reported in literature \cite{Bulanadi2024}.

\section{Conclusion}

In this work, we aimed to clarify the microscopic origins of the setback within recent attempts \cite{Ratzenberger2024} to standardize the preparation of conductive ferroelectric domain walls (DWs) within z-cut monodomain bulk lithium niobate (LN) crystals in terms of reliable final domain shapes and current-voltage (I-V) characteristics. In particular, the DW conductivity 'enhancement' by the application of post-domain-growth voltage ramps resulted in randomly shaped DWs with very different I-V~characteristics so far, even under fully equal process parameters. To investigate possible reasons for this undesired behavior, we conducted an \textit{in-situ} second-harmonic-generation microscopy (SHGM) investigation of the DW evolution during the critical 'enhancement' voltage ramp-up for seven DW samples of different sizes, which allowed us to reconstruct the 3D DW shapes before and after this process, derive the average-inclination changes as well as deducing velocity maps of DW motion along the whole domain perimeter.

As the first main result, we established that vacancy states play a very important role in charge transport in lithium niobate DWs domain walls. It seems that the main role of the conductivity enhancement procedure is not the creation of the DW inclination as described in the literature before, but the attraction of the dopant defect sites, such as ${V_{Li}}$ (p-type) and ${V_O}$ (n-type), since it was shown that the minute inclination of the as-poled domains is already quite enough to maintain high conductance. These results serve as a confirmation of the theoretical prediction made by \citet{Eggestad2024}. Therefore, the point defect accumulation origin of conductance was found to be at least as important as band bending at charged DWs.

An elaborate kinetics analysis of DW movement during the enhancement procedure also implies a large influence of point lattice defects for this process itself -- dopant antisites $Mg_{Li}$, according to the \textit{ab initio} calculation of Lee et al.\cite{Lee2016} -- on their movement (instantaneous velocity, average velocity, mobility at the crystallographically equivalent DW segments, continuity of the movement), and thus, on the final geometrical structure.  The results, outlined here, were already reported for thin film PbTiO$_3$\cite{Yang2024}. It seems that the next step on the way toward reproducibility of both geometrical and conductive properties of DWs in lithium niobate is an artificial tailoring of the point defects' distribution in the crystal (including both position and concentration): the first successful attempts have already been reported in thin films \cite{Bulanadi2024}.

At the present stage, the outlined model for the kinetics of DWs and the conductance mechanism is based on the combination of indirect evidence (3D renders of DWs, their connection with conductance, and the character of the DW motion). For this reason, testing this combined model of defect and h2h/t2t site conduction could benefit from advanced spectroscopic methods, such as electron energy loss spectroscopy, and experimental approaches like patterned ion-beam defect introduction or localized doping near DWs. These methods could isolate and better characterize the contributions of both defects and h2h/t2t sites to LN DW conductance.

\section*{Supplementary Material}

See the supplementary material for a detailed description of SHG image processing, further SHG microscopy images, I-V~data \textit{during} the DW conductivity enhancement voltage ramp-up, standard $\pm$10V~I-V~characteristics, DW inclination atlases for the complete sample set, as well as representations of the DW area dynamics for those samples not shown in the main text.

\section*{Acknowledgments}
We acknowledge financial support by the Deutsche Forschungsgemeinschaft (DFG) through the CRC~1415 (ID: 417590517) and the FOR~5044 (ID: 426703838). This work was supported by the Light Microscopy Facility, a Core Facility of  the CMCB Technology Platform at TU Dresden. I.K.'s contribution to this project is also co-funded by the European Union and co-financed from tax revenues on the basis of the budget adopted by the Saxon State Parliament. I.K. cordially thanks Dr.~Michael Rüsing for encouraging discussions in preparation of this study.

\section*{Author Declarations}

\subsection*{Conflicts of Interest}

The authors have no conflicts to disclose.

\subsection*{Author Contributions}

\textbf{Iuliia Kiseleva:} Conceptualization (equal); Data curation (equal); Formal analysis (equal); Investigation (lead); Methodology (equal); Software (equal); Validation (lead); Visualization (lead); Writing - original draft (lead); Writing –- review \& editing (equal). 
\textbf{Boris Koppitz:} Investigation (equal). 
\textbf{Elke Beyreuther:} Formal analysis (supporting); Funding acquisition (supporting); Supervision (equal); Writing -- review \& editing (lead). \textbf{Matthias Roeper:} Software (equal).
\textbf{Samuel D. Seddon:} Conceptualization (equal); Formal analysis (equal); Methodology (supporting); Supervision (equal); Writing -- review \& editing (equal).
\textbf{Lukas M. Eng:} Funding acquisition (lead); Project administration (lead); Resources (lead); Writing -- review \& editing (equal). 

\section*{Data Availability}

The data that support the findings of this study are available from the corresponding author upon reasonable request.

\section*{References}

\bibliography{main}

\clearpage

\setcounter{table}{0}
\renewcommand{\thetable}{S\arabic{table}}  

\setcounter{page}{1}
\renewcommand{\thepage}{S\arabic{page}} 

\setcounter{equation}{0}
\renewcommand{\theequation}{S.\arabic{equation}} 

\setcounter{figure}{0}
\renewcommand{\thefigure}{S\arabic{figure}}

\setcounter{section}{0}

\onecolumngrid


\section*{SUPPLEMENTARY MATERIAL:}

\subsection{Software for the Characterization of the DW Kinetics and Structure}
\label{SI:sec:software}

In principle, all the acquired grayscale 2D SHGM images of the domains must be transformed into 'skeletons', i.e., binarized images with a one-pixel-wide contour of the DW. The transformation algorithm begins by binarizing the domain wall contour. To account for variations in intensity across different sections of the DW, the contour is divided into smaller regions, each binarized separately using a chosen threshold. Pixels below this threshold are rendered black (0), while those above it are rendered white (1). The resulting domain shape is then reduced to its topological centerline, preserving essential geometric and topological properties, such as connectivity and overall structure. Given the large dataset — hundreds of images for 3D rendering and thousands for time evolution analysis — the process is automated to the highest possible degree.

The process of 3D rendering is realized in four main stages, as shown in Fig.~2 of the main text. The initial stack [Fig.~2(a), main text] of approximately 100 SHG images along the z-axis of the sample is binarized and then stacked with spacing corresponding to the actual physical distance between the acquired scans to form a 3D Point Cloud (PCD) [Fig.~2(b), main text]. Using the Open3D library for Python, a 3D Poisson reconstruction is performed on the PCD to create a triangular mesh. Angles between the domain surface normals and the z-axis are calculated and color-coded according to local inclination [Fig.~2(c), main text]. The 3D model is subsequently 'unwrapped' and presented as a 2D inclination atlas [Fig.~2(d), main text].

For DW kinetics analysis, skeleton areas are calculated as a function of time. To measure local kinetics, the skeleton image is divided into 360 1-degree sectors. For each sector, the distance from the geometric center of the original contour to the sector's edge is extracted for successive time points. These distances are divided by the time interval to obtain \emph{local velocity values}, which are then color-coded in a kinetics map for each of the 360 sectors. To reduce noise and mitigate systematic errors from binarization and skeletonization, a moving average with a window size of 11 points is applied over time to smooth the data.

The following three subsections provide a step-by-step description of the algorithm(s).

\subsubsection{Image scan to skeleton transformation}

The transformation of an SHGM image scan into a skeleton is the foundation for further processing the acquired data. Developing the code for this transformation was the most elaborate part of the entire study, despite the apparent simplicity of the task (after all, the geometrical structures under investigation are fairly simple). These difficulties arose due to the nature of the acquired SHG scans, where the domain wall intensity was often highly inhomogeneous: up to 30\% of the DW signal was barely distinguishable from noise.\\

\paragraph{The basic algorithm} ... for this transformation works as follows:\\

\textit{Step 1:} The areas that do not contain any domain wall structures are masked, including back-poled domains within the primary domain, if present.\\

\textit{Step 2:} The optimal values for noise removal (Gauss kernel and Gauss sigma), i.e., contrast, brightness, dilation iterations, and the initial threshold values, which will be applied to every scan in the stack, are selected manually using sliders for each parameter to provide the cleanest, most distinguishable, and uninterrupted domain wall contour possible. This manual tuning was performed on approximately 5–6 equidistant scans throughout the sample (e.g., layers 1, 20, 40, 60, 80, and 100) for static z-stacks (before and after enhancing DW imaging) or about 30–36 scans for time-dependent stacks (six equidistant time points for six recorded layers). The obtained parameters were averaged over all manually processed images.

\begin{figure*}[htbp]
    \centering
    \includegraphics[width=\textwidth]{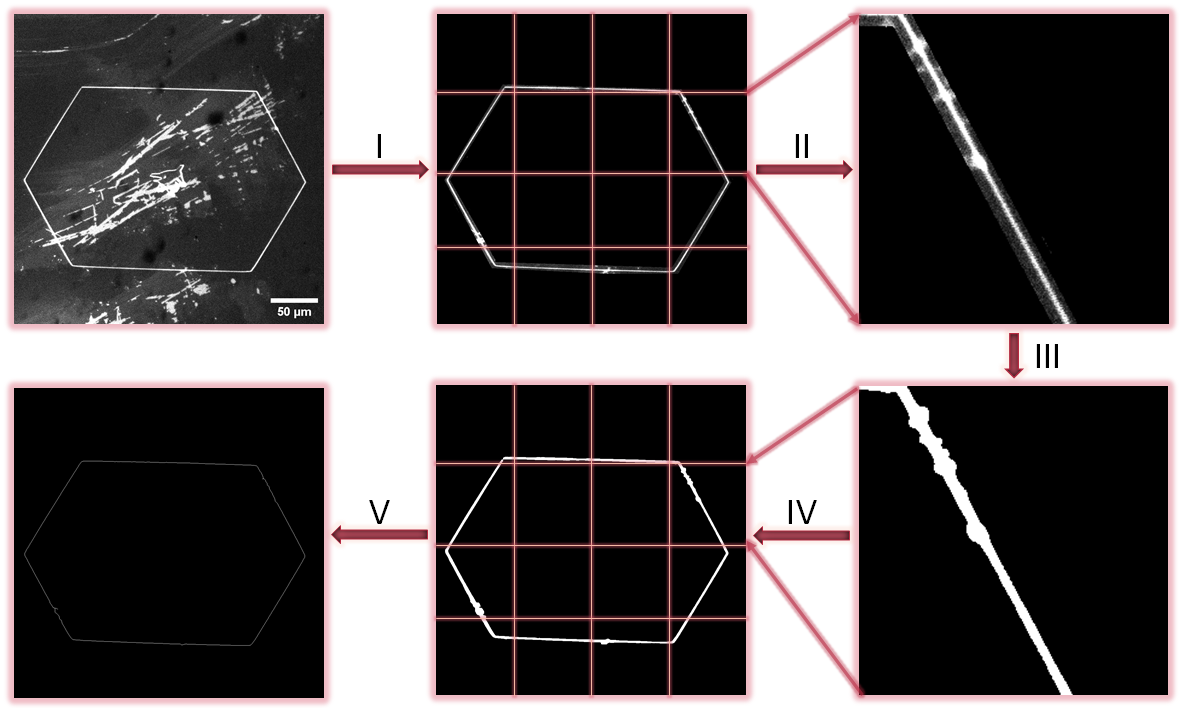}  
    \caption{\justifying \label{fig:binScheme} The step-by-step python algorithm for binarization of the acquired SHG scans. In step (I), the black mask (same for the whole data set) is put on an image. In step (II), the whole image is divided into several pieces, due to the often uneven brightness of the structure across the perimeter. In stage (III), the previous set of parameters is applied to the current piece (noise removal, brightness, contrast, etc.), and the binarization threshold value is automatically adjusted until the contour is uninterrupted. The procedure is repeated for all the pieces, and -- in stage (IV) -- from all the pieces a completely binarized contour is finally received. On stage (V), a command transforms the binarized contour to a one-pixel-thick skeleton.}
\end{figure*}

\textit{Step 3:} The image is divided into pieces both vertically and horizontally. The number of rows and columns must be a power of two (2, 4, 8, etc.), although they do not necessarily have to be equal. The number of divisions depends on the intensity distribution over the entire contour; several variations may need to be tested for each sample. The number of pieces should not be too large (otherwise the segmentation of the picture loses its meaning) or too small (otherwise too many processing defects will be generated); for all the pictures used in this work, both numbers usually took values of 4 or 8.

\textit{Step 4:} Each image piece is modified using the averaged values obtained in Step 2. First, the binarization threshold is set to zero to check whether a domain wall exists in the piece. If white pixels are detected, the threshold is reset to the initial value chosen earlier. Using OpenCV's \texttt{cv.findContours}, the domain wall contour is identified, and \texttt{cv.boundingRect} encloses the detected contour in a rectangle, identifying the rectangle’s corner coordinates. The contour in the image piece is considered closed if at least one of these coordinates lies on the edge of the image piece. If this condition is not met, the binarization threshold is reduced by one, and the continuity of the contour is checked again. This iteration continues until the contour is closed. Then the processed piece is added back to the larger image. The same procedure is applied to all pieces of the larger scan.

\textit{Step 5:} The resulting image contour is then closed (in case of any gaps in the contour) using \texttt{cv.morphologyEx}, and skeletonized with \texttt{morphology.skeletonize} from the SciPy library.

Due to the inhomogeneous distribution of domain wall intensity (with some regions being overexposed and others barely above background noise), this algorithm usually does not yield ready-to-use images for 3D modeling or further calculations. Thus, a second skeletonization step is required.\\

\paragraph{Addressing the three most common defects} ... is the goal of a second step, which treats (a) gaps in the domain wall contour, (b) artifact contours unrelated to the main structure, and (c) branches within the main contour. Often, some combination of these defects is observed. The algorithm should recognize which of these defects are present and correct them in a specific sequence.

The code logic is as follows:

\textit{Step 1:} Here, an initial check for contour closure is performed. All contours in the skeleton are detected, and the main contour (i.e., the longest) is selected. A function is then applied to the main contour to remove loose ends (all non-zero objects are detected; those with one or zero neighbors are identified and removed from the array. The number of iterations is high enough to remove the entire contour if it is not closed.)

\textit{Step 2:} If the number of elements remaining after the first operation exceeds a certain threshold (as a fraction of the original contour length, which accounts for crosses or branches), the contour is considered closed. The same loose-end removal operation is then applied to the entire image, and the result is saved. \textbf{This describes scenario (a): closed contours with possible branches.}

\textit{Step 3:} If almost no elements remain after the initial check, the contour is considered interrupted. If there is only one contour in the image, it is assumed that the contour has only a single break. To eliminate branches, an algorithm identifies the longest path of non-zero elements between loose ends, removing extraneous pixels. The program then draws a straight line between the two open ends, provided the gap is small enough to avoid significant inaccuracies in the SHG signal. \textbf{This describes scenario (b): a contour with one interruption and possible branches.}

\textit{Step 4:} If the initial check detects more than one contour, the domain wall is considered interrupted at multiple points. In this case, each contour fragment is debranched by finding the longest path between its loose ends, removing other extraneous ends. A list of loose-end coordinates is created, and straight lines connect pairs of close coordinates (from different fragments). \textbf{This describes scenario (c): a contour with multiple interruptions and possible excessive branching.}

In the worst case, some layers of the scan are too noisy or contain significant artifacts (e.g., from electrodes, dirt, or surface defects) that interfere with the DW signal. In these cases, the two algorithms described above are insufficient, and the first step of automatic skeletonization must be replaced with manual processing, along with a human-eye comparison between the SHG scan and the processed image. The manual process is similar to the auto-recognition algorithm (which was based on this manual approach), involving masking and breaking the images into equal parts. However, unlike the automatic process, where parameters like Gaussian kernel, sigma, dilation iterations, brightness, and contrast are chosen for a few representative images and averaged over the whole stack, in the manual process, these parameters are adjusted for each part of each layer scan. Additionally, manual adjustment of the binarization threshold is possible, followed by a mandatory post-processing step to remove the most common processing defects, as described above.

\subsubsection{3D Rendering of the DW shape from skeletonized images}

A 3D reconstruction program was necessary to investigate the correlation between the local geometrical structure of the domain wall (specifically, the local inclination angle) and the structure's conductance with a large sample set. The program is fairly simple, using the well-designed Open3D library in Python. This library facilitates 3D imaging with just a few lines of code, provided the input data is well-prepared.

The whole process can be roughly described as depicted in Fig.~2 in the main text. The general algorithm of the code used is as follows: binarized grayscale images (2D arrays) are stacked into 3D arrays, with the stacking distance corresponding to the scan distance between slices. From the non-zero elements, a 3D point cloud (PCD) is formed, and the normals are estimated. In two iterations, triangular meshes are constructed from this PCD. The first triangular mesh from the initial PCD is turned into a second PCD with a more uniform distribution of points. Then, this second PCD is used to generate the final mesh. Normals of the final mesh are properly calculated, and the angles between them and the z-axis of the crystal (which corresponds to the scan axis and is orthogonal to the obtained scans) are computed. Regions of the 3D model are colored according to the calculated angles.

\subsubsection{Calculation of the local domain wall velocity}

Due to the slow image acquisition velocity of the SHGM method, the format of the acquired data (which consists of only six layers inside the sample bulk, with a distance of 30~\textmu m between neighboring scans and a time interval of 0.6--3.5~\text{s} between images of the same layer) allows for local-velocity analyses only within the same 2D layer.

In this work, the local velocity of the domain wall was defined and measured as follows: Under the application of a negative electric field to the +z-side, the domain wall structure's movement was recorded. Throughout the whole process, the larger structures remained convex (protuberant). 

The algorithm for the kinetic analysis is as follows:\\

Calculations are performed in pairs of binarized and skeletonized images of the same layer at adjacent time points, e.g., \textit{layer\_1\_time\_n} and \textit{layer\_1\_time\_n+1}, starting from the first time point. For each image pair, the geometric center coordinates of both contours are found and averaged (to minimize sample movement relative to the camera—if there is no sample movement, the domain wall's movement is smooth enough that the averaged coordinate is essentially the same as the initial centroids). Using these averaged center coordinates, cylindrical coordinates (angle $\varphi$ from 0 to $\pi$ and radius \textit{r}) are assigned to all pixels of both contours.

As mentioned earlier, a one-degree sector of the contour is used as the unit of movement. The algorithm loops through all 360 sectors for both contours; typically, each sector contains 2–4 pixels, depending on the size of the domain wall. The distance from the image center to the contour within each sector is equal to the average \textit{r} coordinate of the pixels in that sector. This distance is calculated for both images in the same loop iteration. The program then computes the difference in distance between these two time points and divides it by the time interval between scans. Positive velocity values indicate inward domain wall movement (as expected during DW conductivity enhancement), while negative values indicate outward movement.

This procedure is repeated for all 360 sectors at time points \textit{n} and \textit{n+1}. The next pair of images at time points \textit{n+1} and \textit{n+2} is then processed equivalently, and the loop continues until the local velocity is calculated for all images. The resulting data is presented as a 3D graph, with the velocity values color-coded and plotted as a function of sector $\varphi$ (x-axis) and time \textit{t} (y-axis). Alternatively, the data can be averaged over time for each of the 360 sectors.

One additional metric for estimating the kinetics of the domain walls is the domain wall area as a function of time and, where applicable, its derivative. Area calculation becomes a simple one-line operation with binarized skeletons available and OpenCV installed.


\subsection{Achieved current enhancement by the DW conductivity enhancement process}

\begin{figure*}[htbp]
    \centering
    \includegraphics[width=\textwidth]{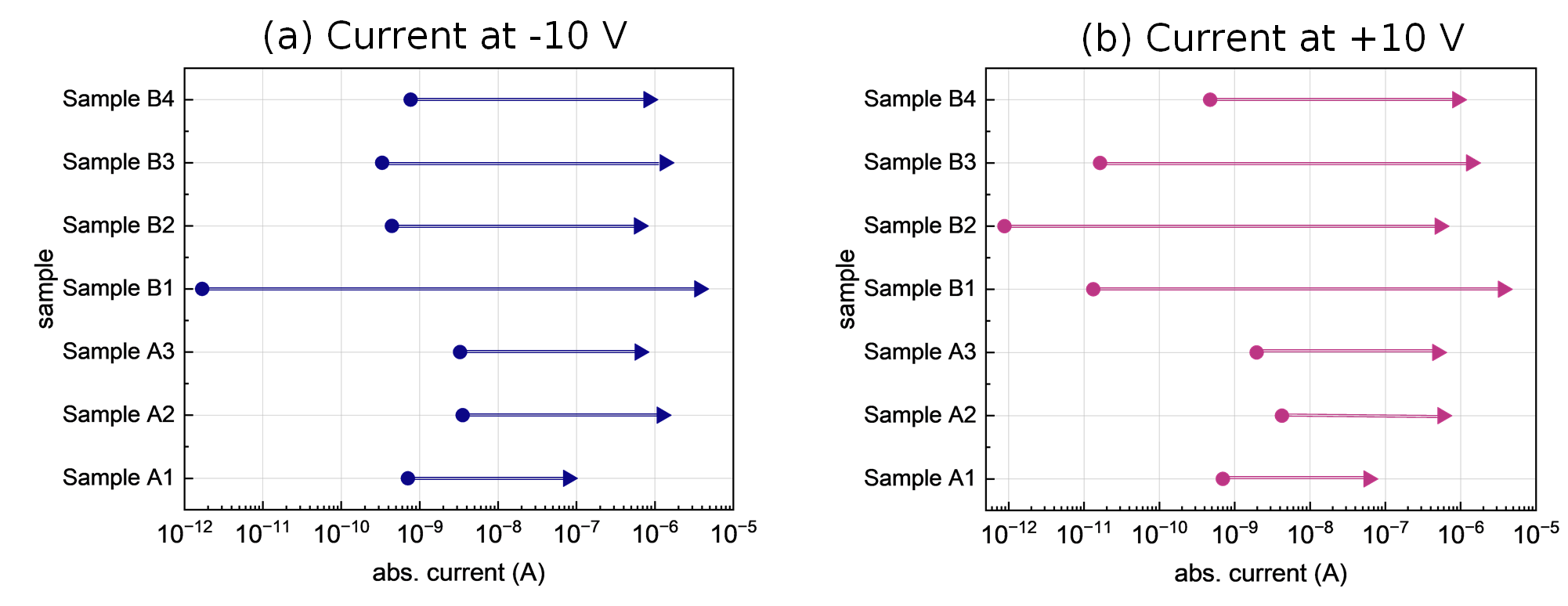}  
    \caption{\justifying \label{fig:Cond_Change}     
    Conductance changes for the investigated samples at -10 V (a) and +10 V (b). Compared to previously reported results \cite{Ratzenberger2024}, a significant improvement in the reproducibility of the enhancement procedure is observed. Despite initial conductivity differences of up to three orders of magnitude in as-poled samples (indicated by circles), 6 out of 7 samples (excluding A1) exhibit currents around 1 \textmu A after enhancement. The mean current at $\pm$10 V is 1.322$\times10^{-6}$~A, with a standard deviation of 1.139$\times10^{-6}$~A. This improvement is attributed to collecting enhancement data across multiple samples, with a limiting current of $10^{-6}$~A  and a stabilization time of 300~s identified as optimal conditions.}
\end{figure*}

\clearpage

\subsection{SHG Images of the Domain Walls in Group B Samples}

\begin{figure*}[htbp]
    \centering
    \includegraphics[scale=0.7,angle=90]{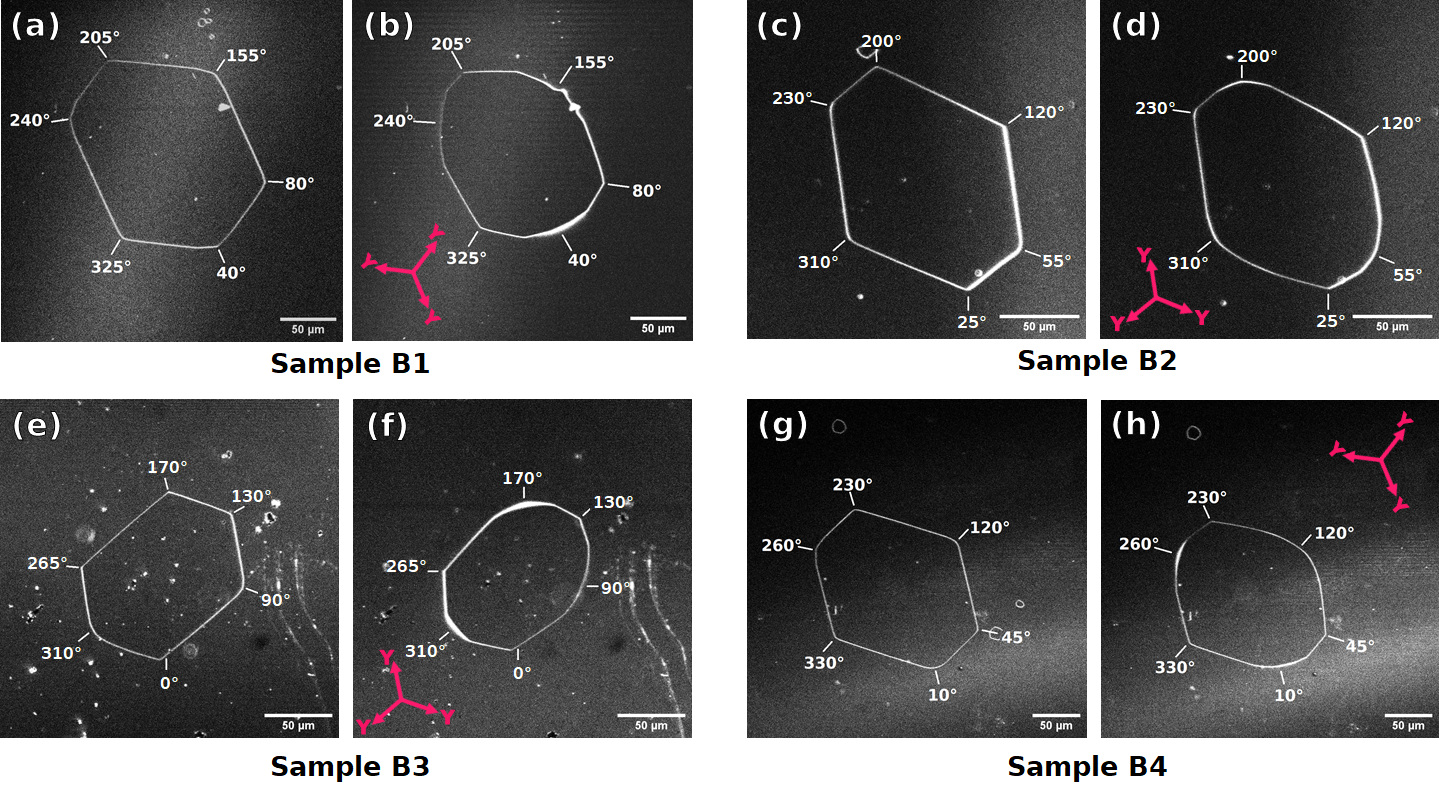}  
    \caption{\justifying \label{fig:SHG_3} SHGM scans of the -z surface before and after enhancement of samples B1 (a, b), B2 (c, d), B3 (e, f), and B4 (g, h) with corresponding $\phi$ angles, used for mapping out inclination and local kinetics of samples.}
\end{figure*}

\clearpage

\subsection{Current-Voltage Characteristics of Group A Samples During the Conductance Enhancement Procedure}

\begin{figure*}[htbp]
    \centering
    \includegraphics[width=\textwidth]{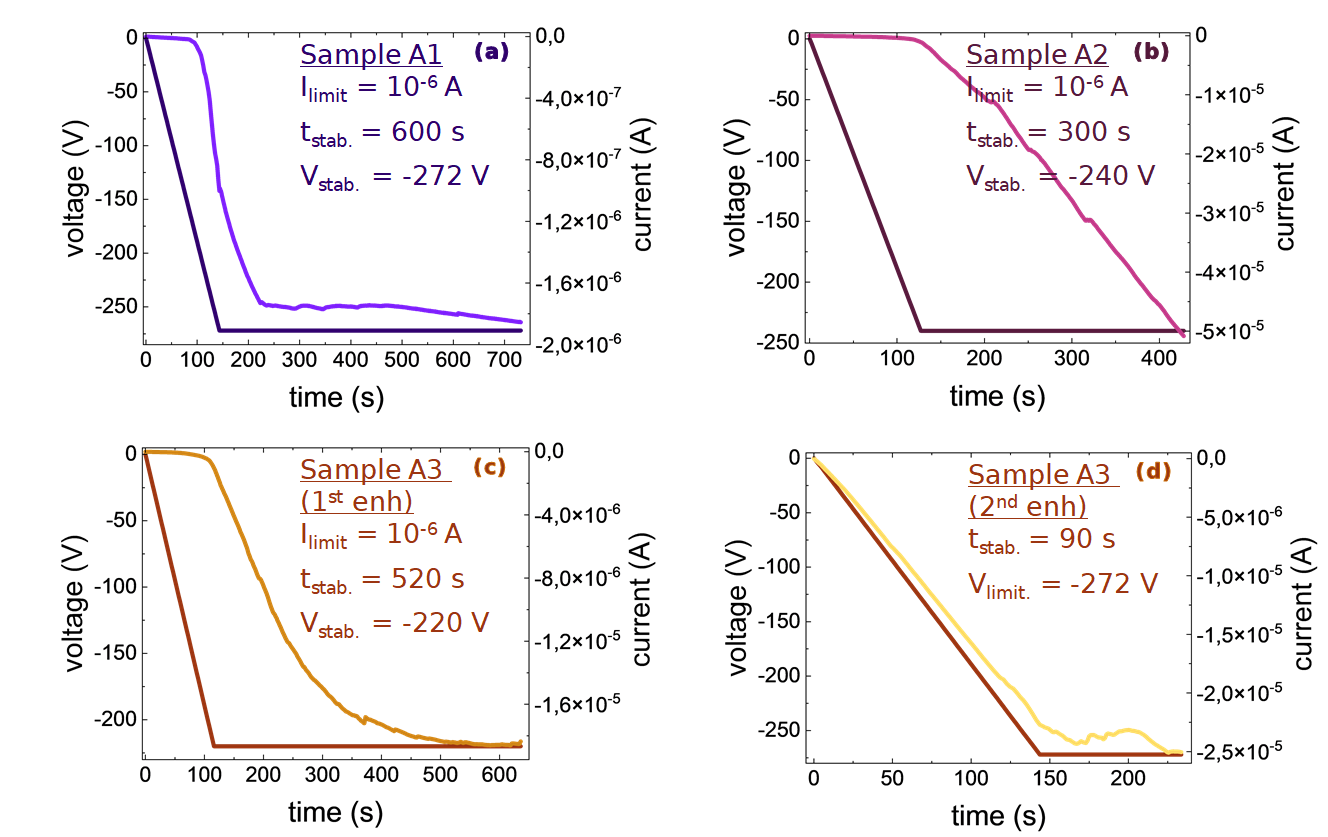}  
    \caption{\justifying \label{fig:Enh_I-V_3} Time dependence of current and voltage during the enhancement procedure for samples A1 (a), A2 (b), and A3 (c) and (d) for two consecutive enhancements. The voltage ramp-up was terminated after the current reached the value of 1 $\mu$ m and was held at this voltage for some more seconds (the former and the latter are specified for each sample on the graph) for further enhancement without implosion. The stabilization was interrupted after no changes to the geometrical structure were detected for a period of time. The second enhancement for sample A3 was conducted in a voltage-limited mode for 90 seconds to avoid implosion. Due to technical reasons, the acquisition of the current for sample A1 has been interrupted after approx. 600 s while application of the voltage was made for 450 s; during this time no significant change of geometry happened, so it is reasonable to assume that the current has remained on the plato.}
\end{figure*}

\clearpage

\subsection{Comparison of Conductance and Geometry Before and After Conductivity Enhancement in Group A Samples}

\begin{figure*}[!htb]
    \centering
    \includegraphics[width=0.875\textwidth]{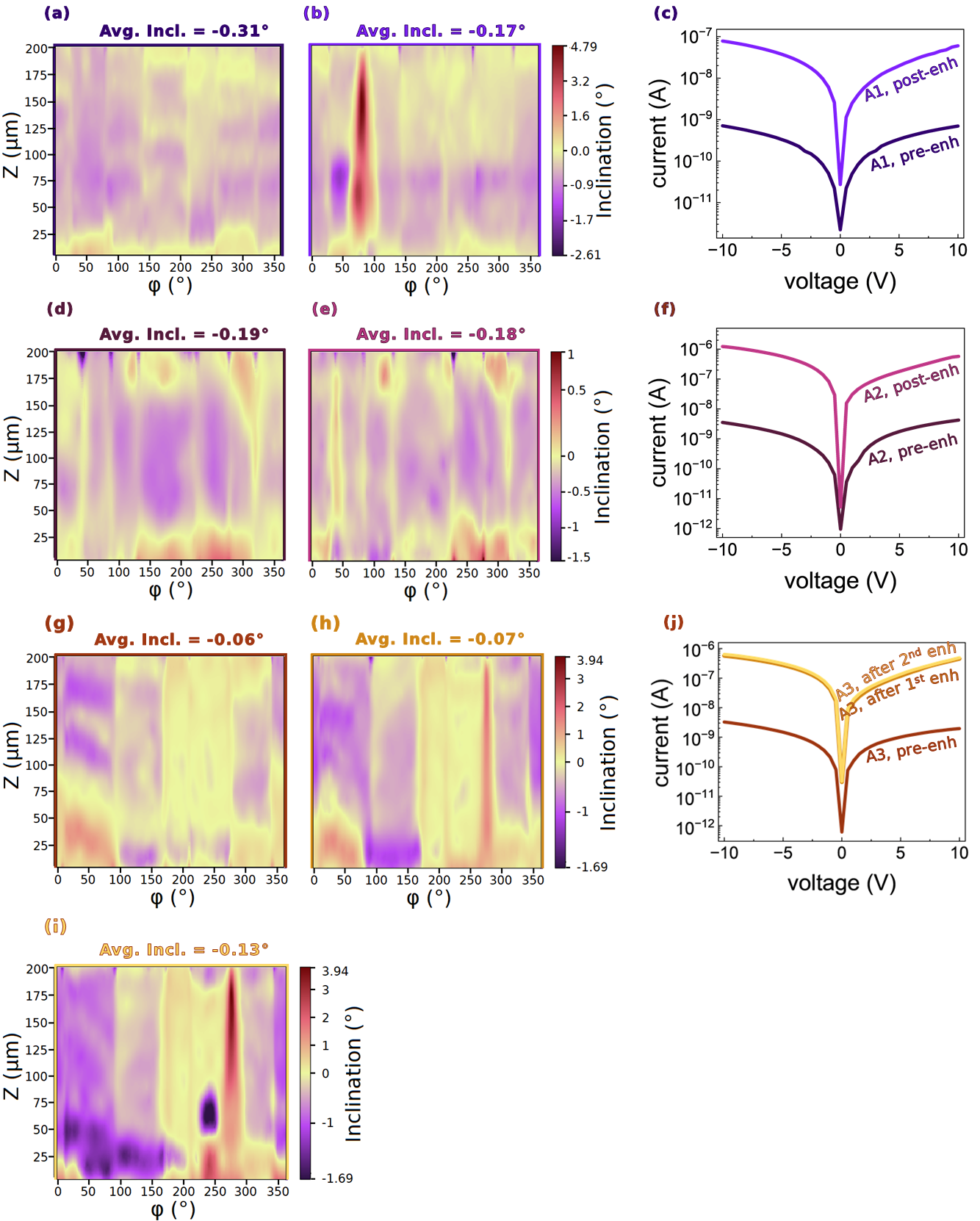}  
    \caption{\justifying \label{fig:InclCond3} Comparison of the geometrical structures before the conduction enhancement - the leftmost inclination maps and after the enhancement - the rightmost inclination maps for samples A1 (the first row), A2 (the second row), A3 (the third and fourth rows) with corresponding current-voltage characteristics before and after the enhancement (graphs c, f, i, and l). This batch with larger domain sizes demonstrates the most conservative changes in geometry during the enhancement process.}
\end{figure*}

\clearpage

\subsection{SHG Images of the Domain Walls in Group B Samples}

\begin{figure*}[htbp]
    \centering
    \includegraphics[scale=0.7,angle=90]{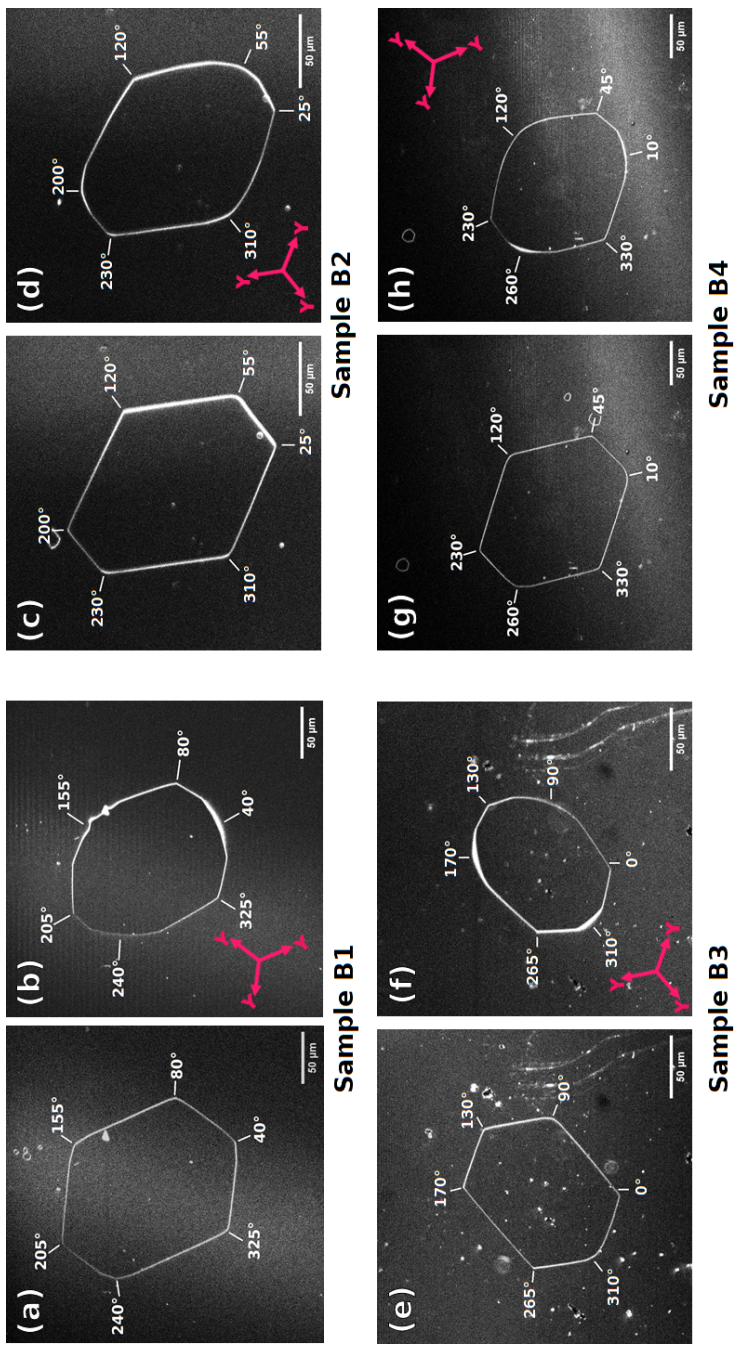}  
    \caption{\justifying \label{fig:SHG_3} SHGM scans of the -z surface before and after enhancement of samples B1 (a, b), B2 (c, d), B3 (e, f), and B4 (g, h) with corresponding $\phi$ angles, used for mapping out inclination and local kinetics of samples.}
\end{figure*}

\clearpage

\subsection{Current-Voltage Characteristics of Group B Samples During the Conductance Enhancement Procedure}

\begin{figure*}[htbp]
    \centering
    \includegraphics[width=\textwidth]{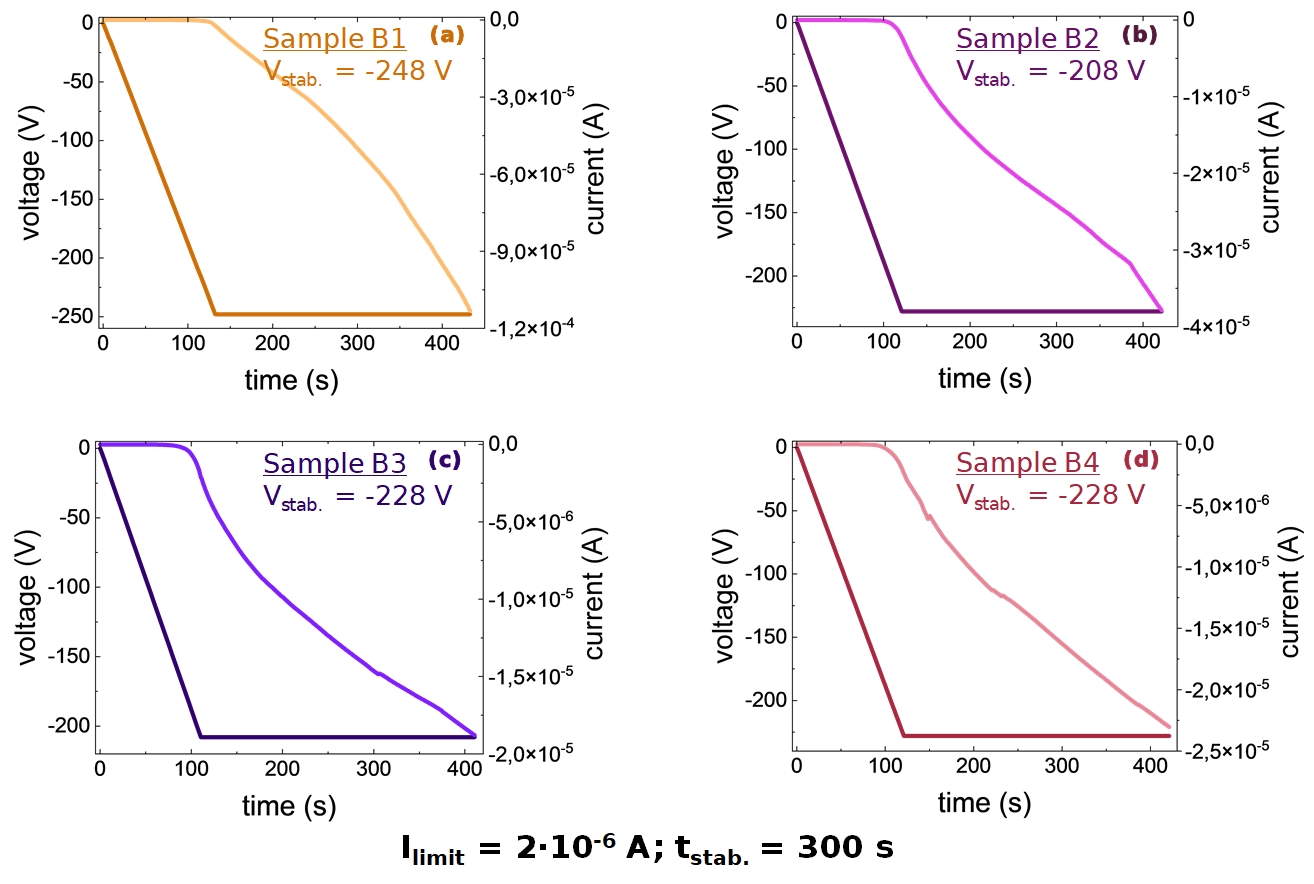}  
    \caption{\justifying \label{fig:Enh_I-V_3} Time dependence of current and voltage during the enhancement procedure for samples B1 (a), B2 (b), B3 (c), and B4 (d). The voltage ramp-up was terminated after the current reached the value of 2 $\mu$ m and was held at this voltage (specified for each sample on the graph) for 300 more seconds for further enhancement without implosion.}
\end{figure*}

\clearpage

\subsection{Comparison of Conductance and Geometry Before and After Conductivity Enhancement in Group B Samples}

\begin{figure*}[!htb]
    \centering
    \includegraphics[width=0.875\textwidth]{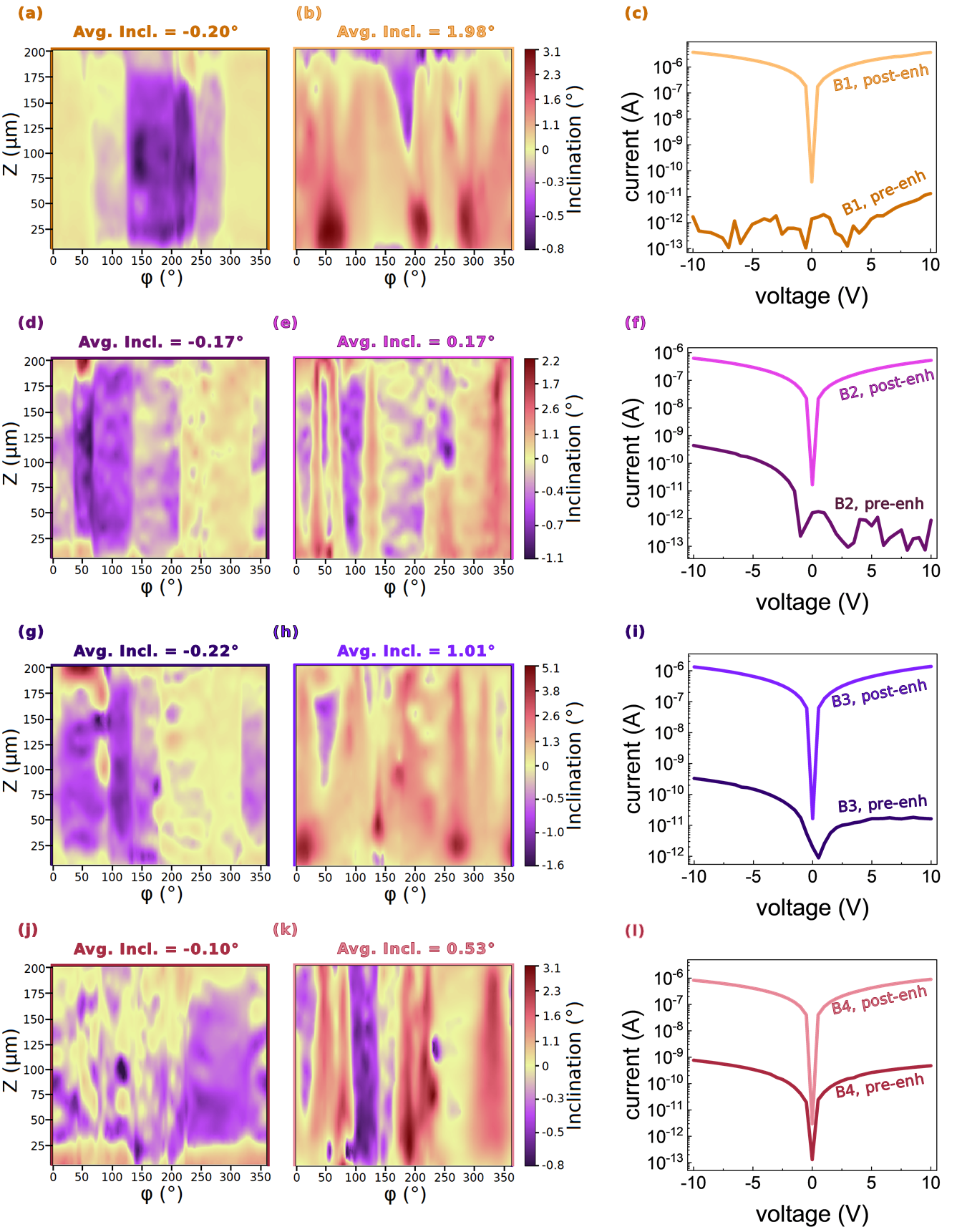}  
    \caption{\justifying \label{fig:InclCond3} Comparison of the geometrical structures before the conduction enhancement - the leftmost inclination maps and after the enhancement - the rightmost inclination maps for samples B1 (the first row), B2 (the second row), B3 (the third row), and sample B4 (the fourth row) with corresponding current-voltage characteristics before and after the enhancement (graphs c, f, i, and l). Inclination maps demonstrate very high roughness of the domain wall planes, serving as a good indicator of the DW geometry dependence on the point defect distribution in the crystal bulk.}
\end{figure*}

\clearpage

\subsection{Area Dynamics Across Six Layers for Samples A2-A3 and B2-B4}

\begin{figure*}[htbp]
    \centering
    \includegraphics[width=0.9\textwidth]{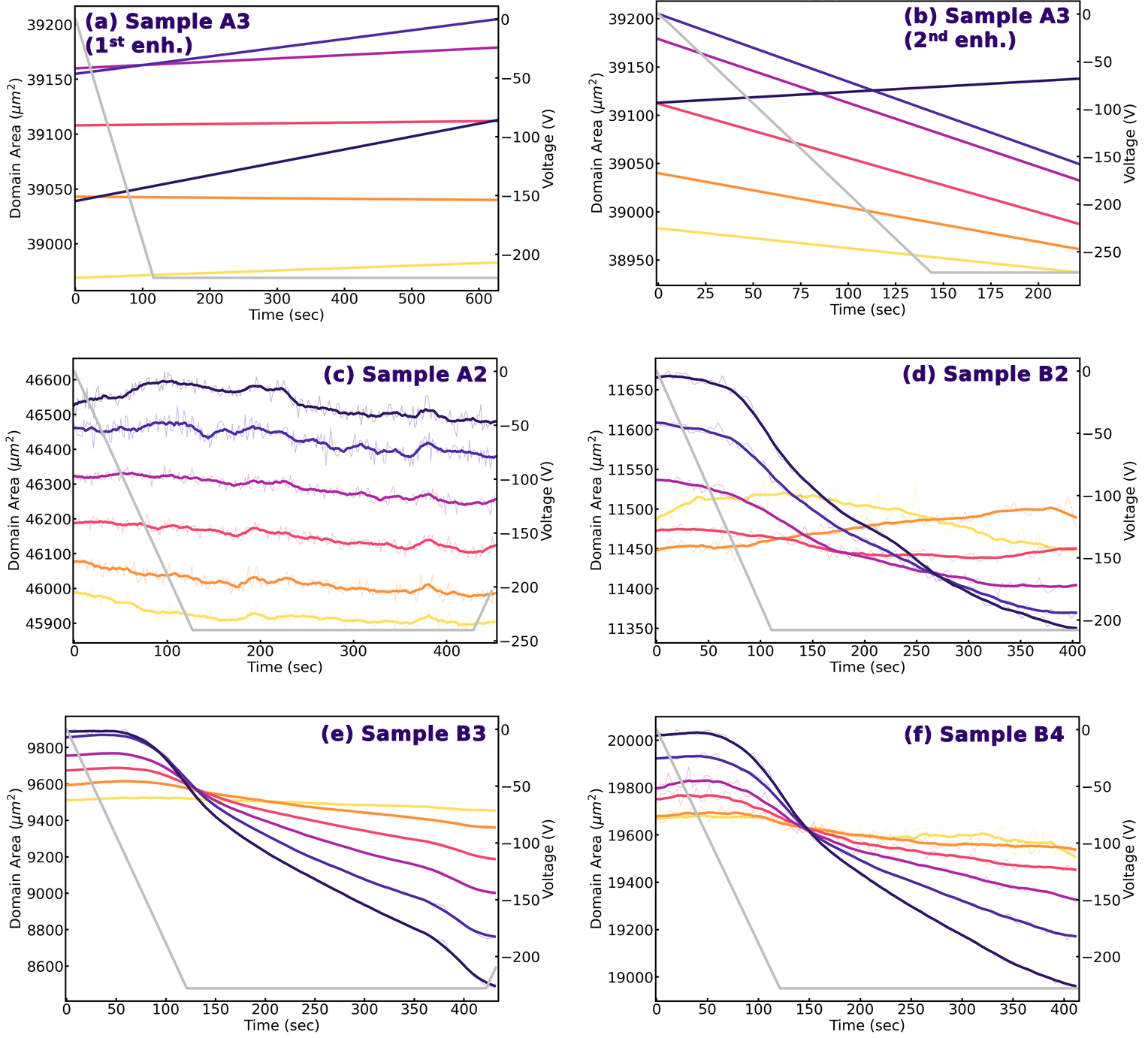}  
    \caption{\justifying \label{fig:Enh_I-V_3} Area dynamics for six equidistantly spaced planes perpendicular to the z-axis of lithium niobate during the enhancement procedure for sample A3 (a, b) - first and second enhancement. There was no in situ acquisition of geometry changes during the enhancement procedure, and the straight lines on the graphs are built from scans along the z-axis before and after enhancement procedures. In general, the averaged dynamics for this sample are very similar to that of group A samples. (c) - domain area dynamics for sample A2 - proof of the absence of the DW dynamics. (d) - sample B2, (e) - sample B3, and (f) - sample B4. All in all, samples of group B have very similar dynamics.}
\end{figure*}

\end{document}